\documentclass[11pt]{article}

\usepackage{amssymb,amsfonts,amsmath,epsfig}
\usepackage[usenames]{xcolor}
\usepackage{times}
\usepackage{pgf,tikz,pgfpages}
\usepackage{authblk}
\usetikzlibrary{arrows,snakes,backgrounds,matrix,topaths,petri,fit,trees,patterns,shapes,shadows}
\usepackage{color,tikz,pgf,pgfarrows,pgfnodes,colortbl,enumerate,multirow}
\usetikzlibrary{plothandlers,plotmarks}
\usepackage{cite}
\usepackage[version=3]{mhchem}
\usepackage{color}
\input xy
\xyoption{all}
\usepackage{setspace}
\usepackage[numbers,square,comma,sort&compress]{natbib}
\newcommand{\R}{\mathbb{R}}

\newcommand{\ie}{\it i.e., \rm}
\newcommand{\eg}{\it e.g., \rm }

 \usepackage{hyperref}

\title{Cellular compartments cause multistability in biochemical reaction networks and allow cells to process more information}

\author[1*]{Heather A. Harrington}
\author[2*]{Elisenda Feliu}
\author[2$\dagger$]{Carsten Wiuf}
\author[13$\dagger$]{Michael P.H. Stumpf}

\affil[1]{\small Theoretical Systems Biology, Division of Molecular Biosciences, Imperial College London, Wolfson Building, London, SW7 2AZ, UK}
\affil[2]{\small Department of Mathematical Sciences, University of Copenhagen, Universitetsparken 5, {2100} Copenhagen, Denmark}
\affil[3]{Institute of Chemical Biology, Imperial College London}
\affil[$\dagger$]{To whom correspondence should be addressed: wiuf@math.ku.dk, m.stumpf@imperial.ac.uk} 
\affil[*]{\small Joint first authors}

\date{\today}

\begin{document}

\maketitle

\begin{abstract}
Many biological, physical, and social interactions have a particular dependence on where they take place. In living cells, protein movement between the nucleus and cytoplasm affects cellular response (i.e., proteins must be present in the nucleus to regulate their target genes). 
Here we use recent developments from dynamical systems and chemical reaction {network} theory to identify and characterize the key-role of the spatial organization of eukaryotic cells in cellular information processing. In particular the existence of distinct compartments  plays a pivotal role in whether a system is capable of multistationarity (multiple response states), and is thus directly linked to the amount of information that the signaling molecules can represent in the nucleus. {Multistationarity provides a mechanism for switching between different response states in cell signaling systems and enables multiple outcomes for cellular-decision making.}
We find that introducing species localization can alter the capacity for multistationarity and mathematically demonstrate that shuttling confers flexibility for and greater control of the emergence of an all-or-none response. 


\end{abstract}

\section*{Introduction}

Cells constantly have to adapt and respond to their environment. In single-celled organisms those cells least well adjusted to their surroundings will tend to contribute less to future generations than cells that are able to assimilate better or more quickly to changing circumstances. In multi-cellular organisms, aberrant response of individual cells to environmental or physiological cues may result in developmental anomalies or disease. The way in which cells respond to external signals, process them and act upon them is thus intimately and inextricably linked to an organism's fate \cite{Cheong:2011jp,Toyoshima:2012hq}; and in the long-term evolution will shape the molecular machinery underlying cellular decision making processes \cite{Balazsi:2011bw,Pearlman:2011eh}.

One central aspect of biological information processing is the mapping of environments onto intra-cellular states {given by} the abundances of the molecular species (proteins, mRNAs, metabolites etc) under consideration. In this processing of information one or more environmental variables need to be represented in a way that facilitates the appropriate response. Continuous and discrete representations have been reported, and it is easy to see how an increased number of states will start to mimic the ``analog" nature of continuously varying states; here, of course, we are typically only interested in stable states.  A simple ``on/off" switch, for example, is a binary representation or response mechanism; this behavior is particularly interesting if there is a regime of conditions where the system can populate either state. In this case we speak of a bistable switch; outside this regime we {only} find a single state for the system. More generally we speak of multistability if more than two {stable} states are obtainable simultaneously. 
\par
The number of states in which a cell can be at any given time is linked to the flexibility in its decision making \cite{Ozbudak:2004gi}. If only one state is accessible (and stable) then there is obviously no room for ``choice" (even if such choices are made by random processes) and any cell-to-cell variability will derive from intrinsic or extrinsic sources of noise which will broaden out the population behavior around such stable states. For bi- and multistable systems, however, cell-to-cell variability may to a large extent be explained in terms of different states being occupied by different cells (even though they are genetically identical). From such variation different cell fates may be differentially accessible and hence understanding the causes of multistability in signal transaction will have ramifications across many areas of modern biology, notably stem-cell biology and  regenerative medicine.
\par
One canonical class of biological systems exhibiting multistability are protein kinase cascades that involve multiple phosphorylation of a substrate \cite{qiao:pcb:2007}. Mitogen activated protein kinase (MAPK) cascades are the most popular exponents of this type of system\cite{ferrell:jbc:1997}. Depending on the mechanism of (de-)phosphorylation bistability in such systems can arise \cite{Thomson:2009fq}, which would give such systems the ability to use different levels of phosphorylation of the final substrate, \eg Erk. This has been an area of growing activity in recent years, because of the important role MAPKs play in cell-fate determination. 
\par
The ultimate function of Erk is to initiate a host of transcriptional responses. To fulfill such a function, activated Erk has to shuttle into the nucleus and a growing body of recent work is paying attention to such spatial aspects of signal transduction \cite{fujioka:jbc:2006,Lidke:2010fr,harrington:physbiol:2012}. Here we show that this spatial organization \cite{kholodenko:nrm:2010} of signal transduction processes plays a pronounced  role in increasing the ``computational space" available to cells. Interestingly, the same effect has been observed at the onset of mitosis \cite{ferrell:cell}.  Very much like the physical {\tt Address space} in a computer processor \cite{FeynmanComp}, the biological equivalent is influenced by the (bio-)physical organization of such systems. And here we show how the compartmentalization increases the number of stable states that can become simultaneously accessible, conferring greater flexibility and plasticity to such systems. Importantly, spatial organization can induce multistability into systems that otherwise would be mono-stable, as well as (sometimes considerably) increase the number of states in systems where the presence of multiple-phosphorylation sites would already give rise to multistable behavior.   
\par
This work complements the findings in \cite{bhalla:biophy}  by providing a detailed mathematical analysis focused on the Erk shuttling mechanisms.
Identifying whether a system exhibits multistable behavior or not, is however, challenging. This type of behavior may be limited to small regions of parameter space and in high-dimensional spaces (our systems considered here have 20 and 36  parameters ({reaction rate constants}), respectively) it is likely not detectable by simulation or random search of the parameter space; instead other approaches are called for. Here we base our arguments on a set of generalizable heuristics that conclusively assert or reject a system's capacity for multistability, and which allow us to identify and delineate multi- and monostable regions in parameter space.

\begin{figure}[t!]
\begin{center}
\includegraphics[scale=0.4]{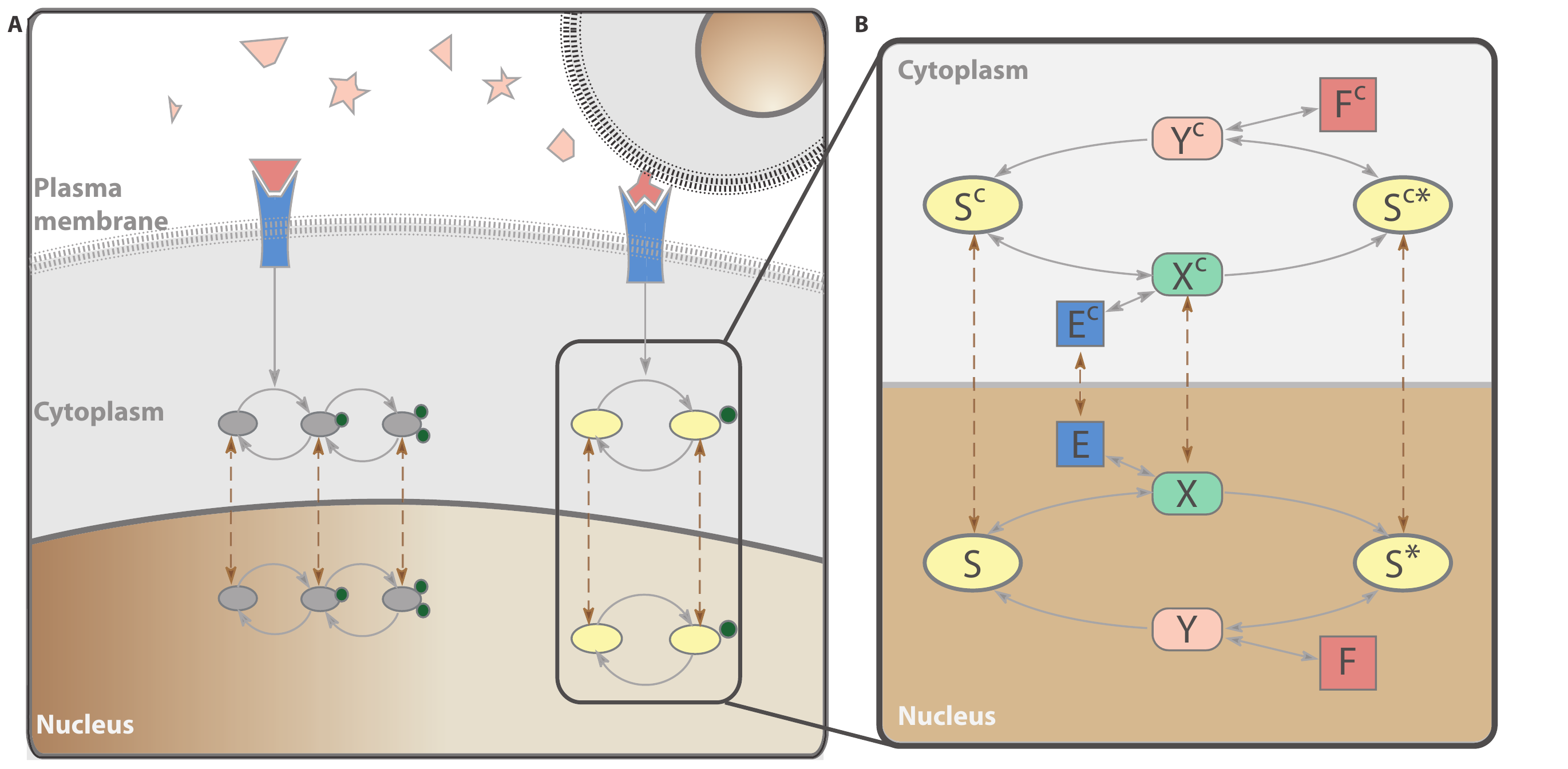}
\caption{\small Spatial signaling schematic. (\textit{A}) Cells require movement of molecular species within the plasma membrane, cytoplasm, nucleus and more cell locations to turn genes on or off and ultimately induce a response. (\textit{B}) One-site phosphorylation/dephosphorylation in two compartments, cytoplasm and nucleus. Molecular species: unphosphorylated substrate ($S$), kinase ($E$), substrate-kinase complex ($X$), phosphorylated substrate ($S^{*}$), phosphatase ($F$), phosphatase- phosphorylated substrate complex ($Y$) and superscript $c$ denotes species in the cytoplasm and all others are in the nuclear compartment. {Kinase  and substrate total abundances ($E_{tot},S_{tot}$) are globally conserved, while phosphatase abundances are conserved within each compartment ($F_{tot},F_{tot}^c$).}}
\end{center}
\end{figure}

\section{Results}
We introduce and illustrate this framework for a basic building block of many signal transduction systems that for spatially homogeneous systems is guaranteed to be monostable. A one-site phosphorylation cycle --- other post-translational modifications can, of course also be considered --- consists of the reversible modification of a substrate $S$ into its phosphorylated form $S^*$.  Phosphorylation and dephosphorylation are  enzymatically catalyzed by $E$ (kinase) and $F$ (phosphatase), respectively, through  a standard Michaelis-Menten mechanism involving the formation of intermediate complexes $X,Y$:
\begin{align} 
E+ S & \cee{<=>[k_{\rm{on},E}][k_{\rm{off},E}]} X   \cee{->[k_{\rm{cat},E}]} E + S^{*} \label{eq:reaction1},\\
F+ S^{*} & \cee{<=>[k_{\rm{on},F}][k_{\rm{off},F}]} Y  \cee{->[k_{\rm{cat},F}]} F + S, \label{eq:reaction2}
\end{align}
where $k_*$ denote the reaction rate constants. 
 Endowed with mass-action kinetics, this  cycle is monostable \cite{goldbeter:pnas:1981,goldbeter:jbc:1984}. 
\par
Now suppose we learn that these enzymatic reactions can both occur in both the nucleus and the cytoplasm (Fig.~1) and we therefore include the reactions in the cytoplasm { (denoted by the $c$ superscript) }
\begin{align}
E^c+ S^c  & \cee{<=>[k_{\rm{on},E}^c][k_{\rm{off},E}^c]} X^c   \cee{->[k_{\rm{cat},E}^c]} E^c + S^{c*}, \label{eq:reaction3}\\
F^c+ S^{c*} & \cee{<=>[k_{\rm{on},F}^c][k_{\rm{off},F}^c]} Y^c  \cee{->[k_{\rm{cat},F}^c]} F^c + S^c, \label{eq:reaction4}
\end{align}
as well as  shuttling reactions between cytoplasm and nucleus \footnote{We remain in a regime where compartmental models are appropriate and where we do not have to model diffusive motion using reaction-diffusion equations. This is appropriate for all cases where transport across a membrane is rate-limiting.}
\begin{align} 
Z^c \cee{<=>[k_{\rm{in},Z}][k_{\rm{out},Z}]} Z, \label{eq:reaction5}
\end{align} 
for the species $Z$ of the one-site {phosphorylation} cycle that shuttle into the nucleus at rate $k_{\rm{in},Z}$ and out at $k_{\rm{out},Z}$. Total kinase, phosphatase and substrate abundances {are} constant for each compartment or globally depending on which species are allowed to shuttle between compartments.  Here we use variation in the amount of active kinase to model how external stimuli are processed and use the substrate state to capture the effects of stimuli.
\par
The species abundances are modeled by a system of ordinary differential equations, which we analyze to determine if the system can exhibit multiple steady states (multistationarity) or not. This analysis employs a suite of different mathematical techniques, including the Jacobian injectivity criterion and recent developments from chemical reaction network theory \cite{Bass:1982p770,pantea-jac,CRNT-toolbox,Feliu-inj}  (see the Appendix for details). When multistationarity can occur we thereby obtain corresponding values of the rate constants and the steady states. Further,  we delimit regions of the parameter space  that contain all sets of {rate constants} that give rise to multistationarity.
\par

\subsection{Analytic conditions for multistability} 
Armed with these tools we  establish that multistationarity cannot occur if only one species shuttles; if two species shuttle, then only the combinations $\{X, S^{*}\}$ or $\{S, Y\}$ can induce multistationarity for certain total amounts and rates; and increasing the number of species that shuttle maintains multistationarity ({see the Appendix for a full description of the sets of shuttling species inducing multistationarity}). The fact that the model includes the formation of {at least one of the} intermediate complexes $X,Y$ (and hence some form of sequestration) is critical for the creation of multistationarity. 
\par
In particular, multistationarity occurs if  the species $E, S, X, S^{*}$ are allowed to shuttle  (Fig.~1B). We choose to study this system in detail  because it {corresponds to a spatial model of a simplified one-site MAPK model system \cite{fujioka:jbc:2006,radhakrishnan:systbiol:2009} that is strictly monostable. }

\begin{figure}[b!]
\begin{center}
\includegraphics[scale=0.4]{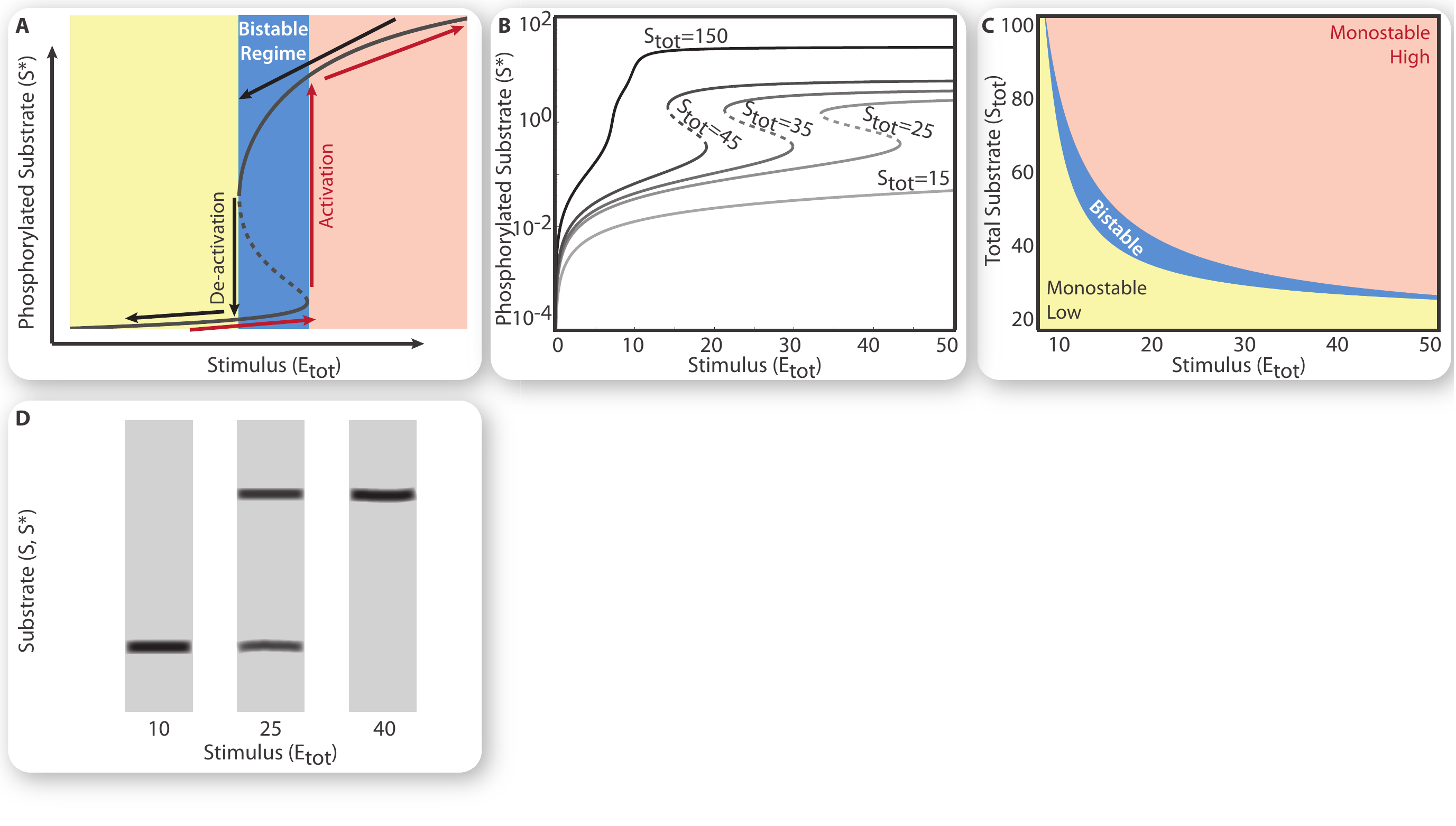}
\caption{\small Bistability in one-cycle localization model. {Rate constants} and total amounts are given in the Appendix. (\textit{A}) Steady state curve of phosphorylated nuclear substrate ($S^{*}$) shows bistability and hysteresis as a function of stimulus (total kinase $E_{tot}$). Stable states, solid lines; unstable state, dashed lines. Activation (phosphorylation) and deactivation (dephosphorylation) switches discontinuously from one stable branch to another at different stimulus thresholds, corresponding to the boundary values of the bistable regime (blue region). (\textit{B}) Steady state curves of $S^{*}$ as a function of stimulus ($E_{tot}$) at varying amounts of total substrate ($S_{tot}$). (\textit{C}) Steady state diagram identifying the regions of parameter space supporting monostability (yellow, orange) or bistability (blue) as a function of the total amounts of kinase ($E_{tot}$) and substrate ($S_{tot}$). (\textit{D}) Theoretical western blot of bistable (two bands) or monostable (one band) behavior depending on dose of stimulus.}
\end{center}
\end{figure}

If the {shuttling rates} fulfill 
\begin{equation}
k_{in,X} \geq k_{\rm{in},E}\quad k_{\rm{out},X}\geq k_{\rm{out},E}\quad k_{\rm{in},S}\geq k_{\rm{in},S^*} \quad k_{\rm{out},S}\geq k_{\rm{out},S^*},
\label{eq:shutrate}
\end{equation}
then multistationarity cannot occur even for the spatial model, whatever total amounts and reaction constants within each cycle (see the Appendix). The shuttling {rates} are paired for species $E,X$ and species $S,S^*$ and therefore a necessary condition for multistationarity is that either $X$ moves in or out  of the nucleus slower than the kinase, $E$; or alternatively that  $S$ shuttles more slowly than its active form, $S^*$.

\begin{figure}[t!]
\begin{center}
\includegraphics[scale=0.4]{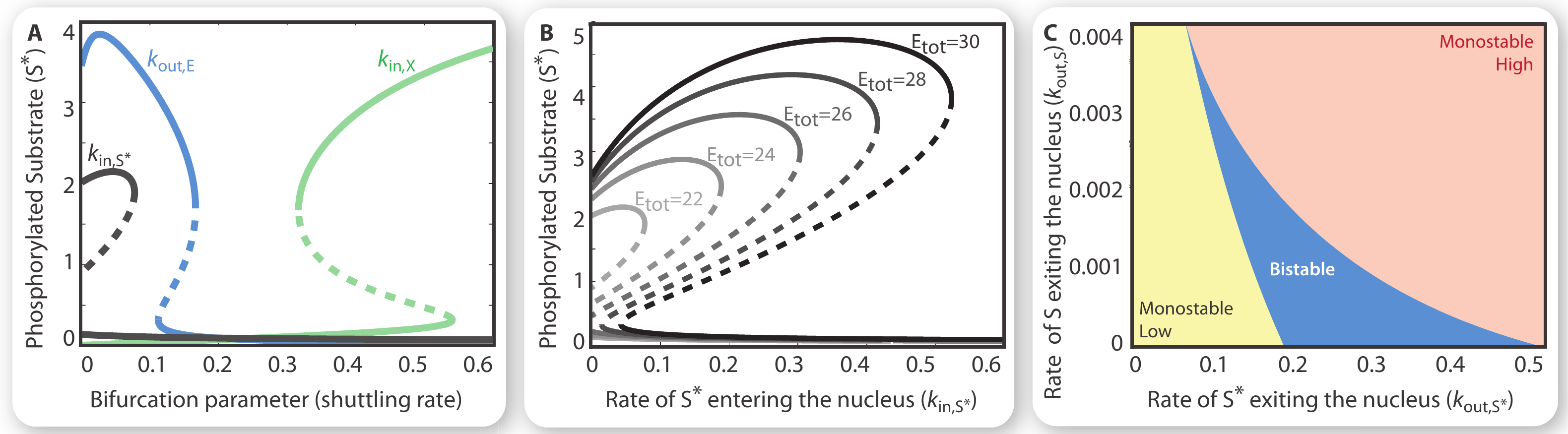}
\caption{\small Effects of shuttling in the one-site localization model. {Rate constants} and total amounts as in Fig.~2.  (\textit{A}) Three steady state curve behaviors of phosphorylated nuclear substrate ($S^*$) as shuttling  are varied: high to low reversible bistability (blue, {$k_{\rm{out},E}$}), low to high reversible bistability (green, {$k_{\rm{in},X}$}), high to low irreversible bistability (black, {$k_{\rm{in},S^*}$}).  (\textit{B}) Steady state curves of phosphorylated nuclear substrate ($S^{*}$) as a function of the {rate constant} of $S^{*}$ shuttling into the nucleus ({$k_{\rm{in},S^*}$}) at varying amounts of stimulus ($E_{tot}$). (\textit{C}) Steady state diagram identifying the regions of parameter space supporting monostability (yellow, orange) or bistability (blue) as a function of {$k_{\rm{out},S}$ and $k_{\rm{out},S^*}$ .} }
\end{center}
\end{figure}

\par
{We next home in on a set of biologically plausible {rate constants} and total abundances for which the system has three steady states, two of which are stable.}  We find that at low and high stimulus doses (\ie total kinase level $E_{tot}=E+E^{c}+X+X^c$) there is one stable steady state of the phosphorylated nuclear substrate, $S^*$, whereas for intermediate stimulus doses two stable steady states coexist (Figs.~2A-C). 
As the stimulus level changes (due to kinase production or degradation), the  state  of $S^{*}$ may switch either to a highly or lowly phosphorylated (activity) steady state based on the system's memory:  switching between states can occur in the form of a hysteresis loop (see black and red arrows in Fig.~2A). 
\par
Shuttling plays a pronounced role in modulating the number of discrete states of the nuclear substrate concentration $S^*$ that can be realized:  an increase of the shuttling {rate constant} causes the system to change either from a high to a low stable state {(for $k_{\rm{out},E},k_{\rm{out},X},k_{\rm{in},S},k_{\rm{in},S^*}$)} or {\em vice versa}  {(for $k_{\rm{in},E},k_{\rm{in},X},k_{\rm{out},S},k_{\rm{out},S^*}$)}, with an unstable steady state in between (Fig.~3A) and  saddle node bifurcations. Depending on which {rate constants} are altered the resulting switches can be  reversible (Fig.~3A, $k_{\rm{out},E},k_{\rm{in},X}$) or irreversible (Fig.~3A, $k_{\rm{in},S^*}$). The response curves of the {rate constants} $k_{\rm{in},E},k_{\rm{in},S^*},k_{\rm{out},X}$, $k_{\rm{out},S}$  can be tuned by altering \eg the total kinase levels, in order to alter the size of multistabile regimes (Fig.~3B) or to turn reversible into irreversible switches. For example, if the shuttling {rate constant}  $k_{\rm{in},S^*}$  is small, then the nuclear $S^*$ can exist in either a high or low state (bistable region); however, following an increase of the shuttling {rate constant} across a threshold,  the system switches to a monostable low state and cannot switch back to the high state nor re-enter the bistable regime. Similarly the {rate constants} $k_{\rm{out},S},k_{\rm{in},E}$ provide irreversible switches favoring the high state. As the shuttling {rate constants} are controlled by a variety of other processes this endows spatially structured systems with a high level of flexibility and increased information-processing ability compared to spatially homogeneous systems: for example,
any violation of the inequalities, Eqns. \eqref{eq:shutrate}, may result in induction of multiple stable states.

\begin{figure}[t!]
\begin{center}
\includegraphics[scale=0.4]{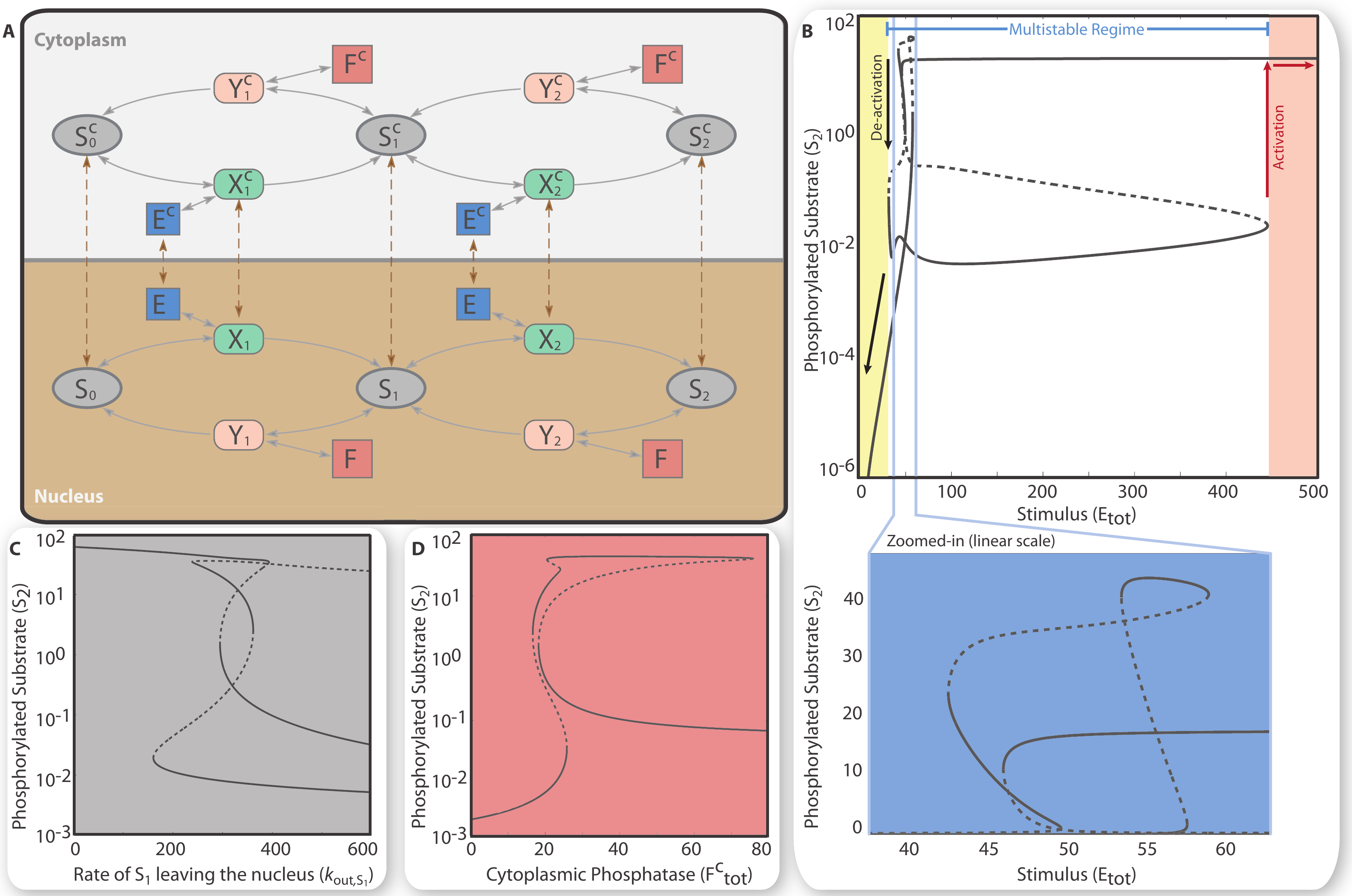}
\caption{\small Two site phosphorylation localization model. {Rate constants} and total amounts provided in the Appendix. (\textit{A}) Schematic of two-site phosphorylation/dephosphorylation in two compartments, cytoplasm and nucleus. (\textit{B}) Steady state curve of phosphorylated nuclear substrate ($S_2$) shows multistability and hysteresis as a function of stimulus (total kinase $E_{tot}$). Stable states, solid lines; unstable states, dashed lines. Activation (phosphorylation) and deactivation (dephosphorylation) switches discontinuously from one stable branch to another at different stimulus thresholds, corresponding to the boundary values of the multistable regime (blue interval). Extreme values of stimulus restrict the system to monostability (log scale). Closer inspection reveals up to four stable states simultaneously (blue figure, zoomed, linear scale). (\textit{C, D}) Steady state curve of phosphorylated nuclear substrate ($S_2$) as a function of the {rate constant} of $S_1$ shuttling out of the nucleus ($k_{\rm{out,S_1}}$) (gray) or as a function of the total cytoplasmic phosphatase ($F_{tot}^{c}$) (coral).}
\end{center}
\end{figure}

\subsection{MAPK two-site phosphorylation}
Having established that species shuttling introduces multistability in a system that otherwise is monostable, we next explore the influence of shuttling  in a system that can already exhibit multistability. Specifically, we consider nuclear localization in a two-site phosphorylation cycle (Fig.~4A), such as the layers of canonical MAPK cascades, and its impact on the cell's ability to establish distinct stable states. Multistability, known to exist in these systems \cite{Markevich-mapk} with up to three stable states \cite{wang-sontag}, has been discussed without reference to any of the spatial models of MAPK signaling   \cite{fujioka:jbc:2006,radhakrishnan:systbiol:2009,shankaran:msb:2009,kholodenko:nrm:2010,kholodenko:wiley:2009}. Here, in order to differentiate between biochemical dependent and shuttling-dependent multistability, we consider biochemical parameter sets that preclude multistability in the absence of localization  (see Appendix). Again spatial structure and shuttling between compartments can induce bistability; \eg fixing the shuttling species to be $E,X_1,X_2,S_0,S_1,S_2$, {bistability is introduced  in the system  (see Appendix)}.
{Provided {that the rate constants} in at least one of the cycles are in the range of multistability of the two-site phosphorylation cycle, we observe that shuttling creates up to 4 stable states and 3 unstable states (Fig.~4B)}. Steady state analysis on a choice of biologically plausible shuttling rates
indicates that {with shuttling} the two-site phosphorylation cycle can undergo hysteresis effects with a large region of multistability ({$32\leq E_{tot} \leq 445$}),  most of which is bistable.
 The doubly-phosphorylated substrate in the nucleus ($S_2$) appears in a low/high monostable state for extremely low/high levels of kinase $E_{tot}$  (red and black arrows in Fig.~4B).
\par
The extended region of multistability with four stable states may provide an explanation for the  versatility of MAPK signaling systems and their widespread use  as relays in many signal transduction networks; it furthermore confers simultaneously both increased robustness and flexibility to the signal processing capabilities of the system compared to spatially homogeneous alternatives.  
As expected, the steady states of the system can be regulated through {\em reversible switches}  governed by shuttling parameters and other total amounts (Figs.~4C, D). But in some cases, stable states may be so close together to be virtually indistinguishable under some physiological conditions; the switching between 3 and 4 stable states in a small region (see zoomed box Fig.~4B) may be an example of this.  This, too, can be modulated, however, by regulation of the shuttling process, or by adjusting total substrate abundances.

\section{Discussion}
Fidelity of information processing and the computational capacity, \ie the ability to map environmental states onto discernible internal states (in particular of proteins active in the nucleus), are thus profoundly affected by the spatial structure of eukaryotic cells. In a biological context, a high and stable state of nuclear substrate ($S^{*}$) can  marshal robust responses to environmental cues. To add further flexibility to such information processing  the shuttling speed of a substrate may further depend on the nature of the stimulus; biological examples abound, and, for example, stimulated NIH 3T3 cells shuttle MAPK into the nucleus three times faster than starved cells \cite{costa:jcs:2006}. By simultaneously controlling the stimulus dose and shuttling speed, a nuanced transition to reversible bistability permits hysteresis, and thus introduces the possibility for switching between low and high stable states (Fig.~3B). 
\par
Trafficking is therefore more intimately related to cellular computation than is typically acknowledged and differences between cell lines in the shuttling {rate constants} of different substrates --- \eg in NIH 3T3 mouse cells, phospho-MAPK can accumulate in the nucleus \cite{costa:jcs:2006}; whereas nuclear accumulation does not occur in PC12 cells \cite{kriegsheim:nat-cell-biol:2009} --- may be hard-wired. Alternatively, the rich set of mechanisms affecting both phosphorylation (exemplary perhaps also for other post-translational modifications) and trafficking give cells the flexibility to change their dynamical regime, \eg from monostable to multistable, on the fly and in response to further environmental and physiological cues. The spatial/compartmental organization of cells thus drives crucial aspects of their information processing capacity; notably the number and robustness of states that can be stably represented is higher (or at least as high) for spatially structured systems compared to homogeneous systems \cite{Cheong:2011jp}. This provides a further rationale for the evolution of cellular compartments \cite{JMSbook} but also begs the question as to how bacteria and archaea can increase their computational capacities. {Here, we believe, micro-environments generated by molecular crowding \cite{McGuffee:2010wl}  confer some of the same advantages. Crowding, \eg around  membrane associated histidine kinases of two-component signaling systems, can isolate signal transduction components spatially from one another, which would have similar (although likely weaker) impact on the computational {\em address space} as cellular compartment have in eukaryotes.}

\bigskip
{\bf Acknowledgments. } The authors gratefully acknowledge funding from The Leverhulme Trust. MPHS is a Royal Society Wolfson Research Merit Award holder. EF is supported by the fellowship ``Beatriu de Pin\'os'' from the Generalitat de Catalunya and the research project MTM2009-14163-C02-01 from Spain. {CW is supported by the Danish Research Council and the Lundbeck Foundation, Denmark.}

\appendix

\section{Appendix: Supporting information}

In the Supporting Information we illustrate the claims made in the main text in more detail. We further expand on the heuristic methods that we have developed to determine if a system of biochemical reactions has the capacity for multiple steady states and to find conditions on the rate constants that ensure that multiple steady states cannot occur. In Section \ref{heuristic} we give an overview of the methods employed.
 In Section \ref{onesite} we study a one-site phosphorylation cycle, which is monostationary, and show that shuttling species can introduce multistationarity. 
In Section \ref{twosite} we study the extended two-site phosphorylation cycle. Without compartmentalization the two-site modification cycle exhibits multistationarity for some choices of rate constants but not all. We show that compartmentalization can introduce multistationarity even if the rate contents do not allow multistationarity in a two-site system without compartmentalization. 

\subsection{Methods for the determination and preclusion of multistationarity}\label{heuristic}
We derive conditions on the rate constants that ensure multistationarity cannot occur. 
Let $$f =(f_1,\ldots,f_n)\colon\R^n\rightarrow \R^n$$ be a differentiable function. The Jacobian, $J_x(f)$, of $f$ at a point $x=(x_1,\ldots,x_n)$ in $ \R^n$  is the $n\times n$ matrix with entry $(i,j)$ being $\frac{\partial f_i}{\partial x_j}$.    If $f$ is a polynomial function then all entries of the matrix $J_x(f)$ are polynomials in $x$ and  consequently,  the determinant of  $J_x(f)$ is a polynomial in $x$ too. 
The Jacobian injectivity criterion states that if  all  components $f_i$ have total degree at most two, and 
the determinant of the Jacobian does not vanish  in a convex domain $\Omega\subseteq \R^n$, then $f$ is an injective function in $\Omega$  \cite{Bass:1982p770}.

The positive steady states of a system of biochemical reactions are given as the positive solutions to a system of polynomial equations $f_{\kappa,i}(x)=0$, $i=1,\ldots,n$, where the coefficients of polynomials $f_{\kappa,i}$ depend on the rate constants $\kappa=\{k_r\}$ and $k_r$ is the rate constant of reaction $r$. If the function $f_{\kappa}=(f_{\kappa,1},\ldots,f_{\kappa,n})$ is injective over the positive real numbers $\R^n_+$  then there can at most be one positive solution to $f_{\kappa}(x)=(0,\ldots,0),$ and consequently,  multistationarity cannot occur. We will use the Jacobian injectivity criterion with $\Omega=\R^n_+$ to determine if the function $f_{\kappa}$ is injective.

In our case, the polynomials $f_{\kappa,i}$ are either quadratic or linear, and hence the Jacobian injectivity criterion can be applied. The determinant of the Jacobian of $f_{\kappa}$ is a polynomial in $x$ with coefficients depending on the rate constants. Each term in the polynomial is of the form $a(\kappa) x_1^{m_1}\cdots x_n^{m_n}$, where $a(\kappa)$ is a coefficient and $m_j$ is 0, 1 or 2. If  the coefficients of the determinant of the Jacobian are all positive or all negative for a specific choice of rate constants, then the determinant cannot vanish when $x$ is positive. Hence, it follows from the Jacobian injectivity criterion that multiple positive steady states cannot occur.  Importantly the coefficients $a(\kappa)$ themselves are  polynomials in the rate constants $\kappa=\{k_r\}$. Thus, we can study how the signs of the coefficients vary when the rate constants vary. Our focus is on  analyzing the coefficients in order to understand what combinations of rate constants make them all  positive or all negative.

The role of the Jacobian injectivity criterion is to preclude multistationarity. However, failure of the Jacobian injectivity criterion is not sufficient to conclude that multistationarity occurs. To investigate whether multistationarity occurs when the Jacobian injectivity criterion fails, we make use of the algorithms implemented in the chemical reaction network theory (CRNT) toolbox \cite{CRNT-toolbox}. For some systems modeled with mass-action kinetics (as is the case here), the toolbox can conclusively determine if multistationarity can occur or not. If the system admits multiple steady states, the toolbox outputs  a unique set of rate constants for which there exists a pair of positive steady states (fulfilling the conservation laws with the same total amounts). However, the rate constants that the toolbox outputs cannot be constrained or controlled in any way. That is to say, we cannot ask the toolbox to restrict  the search to certain regions that are considered biologically realistic. This limits substantially the use of the algorithms of the toolbox. Nevertheless, the output rate constants serve here as a starting point for further investigation and manipulation. 

In this work, a steady state is considered stable if it is asymptotically stable, that is, if the real part of the eigenvalues of the Jacobian of the system at the steady state are all negative. Asymptotic stability ensures that if the initial state of the system is sufficiently close to the steady state then it will eventually be attracted towards the steady state.

\subsection{Shuttling in a one-site phosphorylation cycle}\label{onesite}

\textbf{Reactions and rate constants. }\label{reactions}
We consider a one-site phosphorylation cycle with species $S,S^*$ (the unphosphorylated and phosphorylated substrates), $E$ (kinase), $F$ (phosphatase), and $X,Y$ (intermediate complexes).  Phosphorylation and dephosphorylation are assumed to follow a Michaelis-Menten mechanism (see below and main text). This motif cannot admit multiple steady states and  is monostable \cite{goldbeter:pnas:1981}.
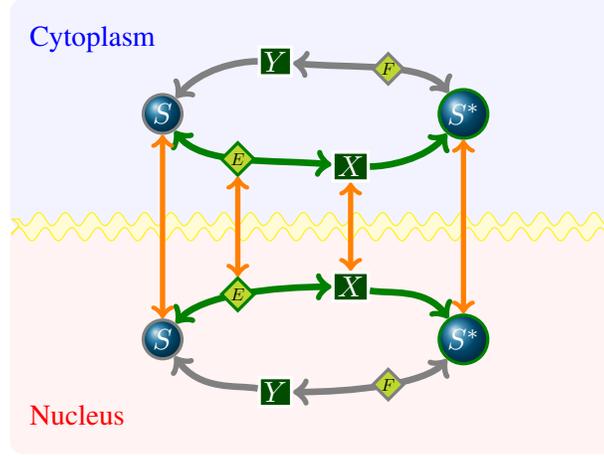
\begin{figure}[!t]
\begin{center}
\begin{tikzpicture}[scale=1]
\node[blue,anchor=west] (s) at (-1.9,5) {Cytoplasm};
\node[red,anchor=west] (s) at (-1.9,0) {Nucleus};
\draw[white,fill=blue, ultra nearly transparent,rounded corners,draw=blue,line width=1.5pt] (-2,2.5) rectangle (6,5.5);

\draw[white,fill=red, ultra nearly transparent,rounded corners,draw=red,line width=1.5pt] (-2,2.5) rectangle (6,-0.5);
\draw[fill=yellow!20!white,draw=yellow,snake=coil,segment aspect=0] (-2,2.6) rectangle (6,2.4);
\node[white,circle,ball color=blue!60!green,draw=gray,line width=1.2pt,inner sep=1.5pt] (S11) at (0,4)  {$S$};
\node[white,circle,ball color=blue!60!green,draw=green!50!black,line width=1.2pt,inner sep=1.5pt] (S21) at (4,4)  {$S^*$};

\node[white,circle,ball color=blue!60!green,draw=gray,line width=1.2pt,inner sep=1.5pt] (S112) at (0,1)  {$S$};
\node[white,circle,ball color=blue!60!green,draw=green!50!black,line width=1.2pt,inner sep=1.5pt] (S212) at (4,1)  {$S^*$};
\node[white,rectangle,draw=white,fill=green!30!black,line width=1.2pt,inner sep=1.7pt] (X) at (2.5,3.3)  {$X$};
\node[white,rectangle,draw=white,fill=green!30!black,line width=1.2pt,inner sep=1.7pt] (Y) at (1.5,4.7)  {$Y$};

\node[white,rectangle,draw=white,fill=green!30!black,line width=1.2pt,inner sep=1.7pt] (X2) at (2.5,1.7)  {$X$};
\node[white,rectangle,draw=white,fill=green!30!black,line width=1.2pt,inner sep=1.7pt] (Y2) at (1.5,0.3)  {$Y$};

\draw[<->,draw=green!50!black,line width=2.5pt] (S11) .. controls (0.5,3.3) and (1.5,3.4) .. (X);
\draw[->,draw=green!50!black,line width=2.5pt] (X) .. controls (3,3.3) and (3.5,3.4) .. (S21);
\draw[<->,draw=gray,line width=2.5pt] (S21) .. controls (3.5,4.7) and (2.5,4.6) .. (Y);
\draw[->,draw=gray,line width=2.5pt] (Y) .. controls (1,4.7) and (0.5,4.6) .. (S11);

\draw[<->,draw=green!50!black,line width=2.5pt] (S112) .. controls (0.5,1.6) and (1.5,1.7) .. (X2);
\draw[->,draw=green!50!black,line width=2.5pt] (X2) .. controls (3,1.6) and (3.5,1.7) .. (S212);
\draw[<->,draw=gray,line width=2.5pt] (S212) .. controls (3.5,0.4) and (2.5,0.3) .. (Y2);
\draw[->,draw=gray,line width=2.5pt] (Y2) .. controls (1,0.4) and (0.5,0.3) .. (S112);

\node[diamond,draw=green!50!black,line width=1.2pt,fill=yellow!70!green,scale=0.6,inner sep=1.2pt] (E2) at (1,1.6)  {$E$};
\node[diamond,draw=gray,line width=1.2pt,fill=yellow!70!green,scale=0.6,inner sep=0.9pt] (F2) at (3,0.4)  {$F$};

\node[diamond,draw=green!50!black,line width=1.2pt,fill=yellow!70!green,scale=0.6,inner sep=1.2pt] (E) at (1,3.4)  {$E$};
\node[diamond,draw=gray,line width=1.2pt,fill=yellow!70!green,scale=0.6,inner sep=0.9pt] (F) at (3,4.6)  {$F$};

\draw[<->,orange,line width=2pt] (E) -- (E2);
\draw[<->,orange,line width=2pt] (X) -- (X2);
\draw[<->,orange,line width=2pt] (S11) -- (S112);
\draw[<->,orange,line width=2pt] (S21) -- (S212);

\end{tikzpicture}
\end{center}
\caption{Shuttling  of a one-site phosphorylation cycle between the nucleus and the cytoplasm.}\label{Fig:1}
\end{figure}

To study the effect of compartmentalization we assume that the species $S,S^*,E,X$ can shuttle between the cytoplasm and the nucleus (see Figure~\ref{Fig:1}). We let $Z^c$  denote the species $Z$ in the cytoplasm. Then, we have the following reactions:

\begin{itemize}
\item Reactions in the nucleus:

 \centerline{\xymatrix{
E + S \ar@<0.3ex>[r]^(.6){k_1}  & X  \ar@<0.3ex>[l]^(.4){k_2}  \ar[r]^(.4){k_3} & E + S^*  &
F + S^* \ar@<0.3ex>[r]^(.6){k_4}  & Y  \ar@<0.3ex>[l]^(.4){k_5}  \ar[r]^(.4){k_6}  & F+S
}}

\item Reactions in the cytoplasm:

 \centerline{\xymatrix{
E^c + S^c \ar@<0.3ex>[r]^(.6){k_7}  & X^c  \ar@<0.3ex>[l]^(.4){k_8}  \ar[r]^(.4){k_9} & E^c + S^c  &
F^c + S^{c*} \ar@<0.3ex>[r]^(.6){k_{10}}  & Y^c  \ar@<0.3ex>[l]^(.4){k_{11}}  \ar[r]^(.4){k_{12}}  & F^c+S^{c*}
}}

\item Shuttling reactions:

 \centerline{\xymatrix{
E  \ar@<0.3ex>[r]^(.5){k_{13}}  & E^c  \ar@<0.3ex>[l]^(.5){k_{17}}  &  X  \ar@<0.3ex>[r]^(.5){k_{14}}  & X^c  \ar@<0.3ex>[l]^(.5){k_{18}} & S  \ar@<0.3ex>[r]^(.5){k_{15}}  & S^c  \ar@<0.3ex>[l]^(.5){k_{19}}  & S^*  \ar@<0.3ex>[r]^(.5){k_{16}}  & S^{c*}  \ar@<0.3ex>[l]^(.5){k_{20}}   
}}
\end{itemize}

To ease the notation, we have changed the notation of the reaction constants $k_r$ in the main text and simply labeled them with consecutive numbers $k_1,\dots,k_{20}$. The  correspondence between the  two notations is shown below:

\begin{center}
\begin{tabular}{|c|c|c|c|c|c|c|c|c|c|c|c|c|c|c|c|c|c|c|c|c|} \hline
\rowcolor{lightgray}
Here & $k_1$ &  $k_2$ &  $k_3$ &  $k_4$ & $k_5$ &  $k_6$ &  $k_7$ &  $k_8$ &$k_9$ &  $k_{10}$  \\ 
Main text &  $k_{\rm{on},E}$ & $k_{\rm{off},E} $& $ k_{\rm{cat},E}$ & $ k_{\rm{on},F}$ &$ k_{\rm{off},F}$ &  $k_{\rm{cat},F}$ &  $k_{\rm{on},E}^c$   & $k_{\rm{off},E}^c$ & $k_{\rm{cat},E}^c$ & $k_{\rm{on},F}^c$\\ \hline \hline
\rowcolor{lightgray}
Here & $k_{11}$ &  $k_{12}$ &$k_{13}$ &  $k_{14}$ &  $k_{15}$ &  $k_{16}$  & $k_{17}$ &  $k_{18}$ &  $k_{19}$ &  $k_{20}$\\
Main text &   $ k_{\rm{off},F}^c$ &  $k_{\rm{cat},F}^c$ & $k_{\rm{out},E}$
 & $k_{\rm{out},X}$ & $k_{\rm{out},S}$ & $k_{\rm{out},S^*}$ & $k_{\rm{in},E}$
 & $k_{\rm{in},X}$ & $k_{\rm{in},S}$ & $k_{\rm{in},S^*}$ \\ \hline
 \end{tabular}
\end{center}

\medskip
\noindent
\textbf{Mass-action system of ordinary differential equations. }\label{equations}
We order the set of species in the following way:
$$(x_1,x_2,x_3,x_4,x_5,x_6)=(E, X, S, S^*, F, Y),\quad (x_7,x_8,x_9,x_{10},x_{11},x_{12})=(E^c, X^c, S^c, S^{c*}, F^c, Y^c).$$
By assuming the law of mass-action, the dynamics of this reaction network  is modeled by the following system of ordinary differential equations (reference to  time $t$ is omitted, $x_i=x_i(t)$):
\begin{align}
\dot{x}_1 &=-k_{13} x_{1} + k_{2} x_{2} + k_{3} x_{2} - k_{1} x_{1} x_{3} + k_{17} x_{7}, \nonumber\\ 
\dot{x}_2  &=  -k_{2} x_{2} - k_{3} x_{2} - k_{14} x_{2} + k_{1} x_{1} x_{3} + k_{18} x_{8},\nonumber\\ 
\dot{x}_3 &=  k_{2} x_{2} - k_{15} x_{3} - k_{1} x_{1} x_{3} + k_{6} x_{6} + k_{19} x_{9},\nonumber\\ 
\dot{x}_4 &=  k_{3} x_{2} - k_{16} x_{4} - k_{4} x_{4} x_{5} + k_{5} x_{6} + k_{20} x_{10},\nonumber\\ 
\dot{x}_5 &=  -k_{4} x_{4} x_{5} + k_{5} x_{6} + k_{6} x_{6},\nonumber\\ 
\dot{x}_6 &= k_{4} x_{4} x_{5} - k_{5} x_{6} - k_{6} x_{6},\label{odes}\\ 
\dot{x}_7 &=  k_{13} x_{1} - k_{17} x_{7} + k_{8} x_{8} + k_{9} x_{8} - k_{7} x_{7} x_{9}, \nonumber \\ 
\dot{x}_8 &=  k_{14} x_{2} - k_{8} x_{8} - k_{9} x_{8} - k_{18} x_{8} + k_{7} x_{7} x_{9}, \nonumber\\ 
\dot{x}_9 &=  k_{15} x_{3} + k_{8} x_{8} - k_{19} x_{9} - k_{7} x_{7} x_{9} + k_{12} x_{12}, \nonumber \\ 
   \dot{x}_{10} &=  k_{16} x_{4} + k_{9} x_{8} - k_{20} x_{10} - k_{10} x_{10} x_{11} +  k_{11} x_{12}, \nonumber \\ 
\dot{x}_{11} &= -k_{10} x_{10} x_{11} + k_{11} x_{12} + k_{12} x_{12}, \nonumber \\ 
 \dot{x}_{12} &= k_{10} x_{10} x_{11} - k_{11} x_{12} - k_{12} x_{12}.\nonumber
\end{align}

This dynamical system has four conservation laws, accounting for the fact that the amounts of enzymes and substrate are conserved:
\begin{align}
0 &= \dot{x}_1+\dot{x}_2+\dot{x}_7+\dot{x}_8, \nonumber\\ 
0 &=\dot{x}_5+\dot{x}_6, \label{conseq} \\ 
0 &= \dot{x}_2+\dot{x}_3+\dot{x}_4+\dot{x}_6+\dot{x}_8+\dot{x}_9+\dot{x}_{10}+\dot{x}_{12}, \nonumber\\ 
0 &=\dot{x}_{11}+\dot{x}_{12}.\nonumber
\end{align}
These conservation laws can be verified but adding the corresponding equations in \eqref{odes}. Since the model does not incorporate shuttling of the phosphatase, the amount of phosphatase is conserved separately  in each compartment.

Let $S_{tot}$ denote the total amount of substrate, and $E_{tot},F_{tot},F_{tot}^c$ denote the total amounts of kinase and phosphatase in the system, respectively.  The differential equations in \eqref{conseq} lead to the following equations that are fulfilled at  any time:
\begin{align}
E_{tot} &=   x_1+x_2+x_7+x_8  \nonumber \\ 
  F_{tot} &= x_5+x_6 \label{totamounts} \\
S_{tot} &= x_2+x_3+x_4+x_6+x_8+x_9+x_{10}+x_{12} \nonumber \\
 F_{tot}^c &= x_{11}+x_{12}. \nonumber
 \end{align}

The steady states of the system are obtained by setting all derivatives $\dot{x}_i$ to zero.
The system has the \emph{capacity for multiple steady states} if there exist rate constants $k_1,\ldots,k_{20}$ and positive total amounts $S_{tot},E_{tot},F_{tot},F_{tot}^c$ such that the equations
$\dot{x_i}=0$ together with \eqref{totamounts} have more than one positive solution. Therefore, for fixed reaction rates and total amounts,  determination of multistationarity implies  solving a system of polynomial equations in 12 indeterminates (variables). The equations corresponding to the conservation laws are linear, while those corresponding to setting the derivatives to zero are quadratic (that is, they have terms of total degree 1 and 2).

\newpage
\noindent
\textbf{Rate constants and total amounts for multistationarity (for Figs. 2, 3 in the main text). }\label{ratesmulti}
The CRNT toolbox provides a unique set of rate constants  for which the system admits multiple positive steady states
{\small \begin{align*}
k_1& =11.679195 & k_2&= 144.94137  & k_3&=91.527059 & k_4&= 207.26904 & k_5&= 22.115015, \\ 
 k_6&= 309.97808,  & k_7&=49.545796, & k_8&=8.8750284, & k_{9}&=262.90818, & k_{10}&=356.03934, \\
 k_{11}&= 1.8978202, & k_{12}&=44.457164, & k_{13}&=  1.0903408, & k_{14}&=305.42214, & k_{15}&= 47.547732, \\
 k_{16}&=41.866754, & k_{17}&=86.473107, & k_{18}&=215.67801, & k_{19}&=1,& k_{20}&=  165.98446.
   \end{align*}}
For this set of rate constants, two steady states are provided  with total amounts:
$$ E_{tot}= 20.7066814, \quad S_{tot} =35.21053215, \quad F_{tot} = 3.84921092, \quad F_{tot}^c= 11.0903086.$$

We aim to exemplify multistationarity with rate constants that are more biologically reasonable and of the order of experimentally determined values \cite{fujioka:jbc:2006,harrington:physbiol:2012}.
To this end, we have manually investigated the effect of changing a specific rate or a total amount with respect to the  emergence of multistationarity. We guide the proposed changes by the structure of the steady-state equations.
This procedure has allowed us to tune the  rate constants and total amounts to reasonable values without loosing multistationarity. Specifically, we settled for the rates (used to create Figures 2 and 3 in the main text):
{\small \begin{align*}
k_1 &= 0.049 & k_2 &= 0.009 & k_3 &= 0.262 & k_4 &= 0.356 & k_5 &= 0.002 & k_6 &= 0.044 & k_7 &= 0.011
\\  k_8 &= 0.144 & k_9 &= 0.091 & k_{10} &= 0.207 & k_{11} &=0.022 & k_{12} &= 0.309 & k_{13} &= 0.16 & k_{14} &=  0.14 \\ k_{15} &=  0.001 & k_{16} &= 0.166 & k_{17} &= 0.0006 & k_{18} &= 0.33 & k_{19} &= 0.047 & k_{20} &= 0.041,
\end{align*}}
and the total amounts
$\{E_{tot},S_{tot},F_{tot},F^c_{tot}\}=\{22, 35, 11, 3\}$. 
With these parameters, there are three steady states, of which two are stable.  Specifically, the  steady states are approximately:
{\small \begin{align*}
SS_1 &= (0.676,2.357,16.33,1.261,1.023,9.977,17.671,1.296,2.068,0.751,2.041,0.959) \\ 
  SS_2 &= (0.183,0.965,27.220,0.126,5.623,5.377,20.391,0.461,0.559,0.107,2.812,0.188) \\
  SS_3 &= (1.062,2.87,11.847,2.145,0.625,10.375,16.371,1.701,3.109,1.503,1.546,1.454).
  \end{align*}}
  The steady state $SS_1$ is unstable and has only one eigenvalue with positive real part. 

\medskip
\noindent
\textbf{Conditions for monostationarity (Eqns (6) in the main text). }\label{inequalities}
Not all choices of rate constants and total amounts have the capacity for multistationarity. We show here that there is a set of  \emph{necessary} conditions for the existence of multistationarity that depends exclusively on the shuttling rates.
To see this, we apply the Jacobian injectivity criterion to a function that in part consists of  the right hand sides of the conservation laws \eqref{odes}.

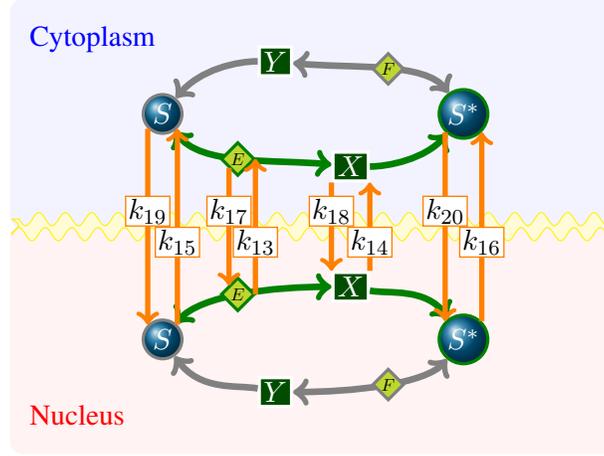
\begin{figure}[t]
\begin{center}
\begin{tikzpicture}[scale=1]
\node[blue,anchor=west] (s) at (-1.9,5) {Cytoplasm};
\node[red,anchor=west] (s) at (-1.9,0) {Nucleus};
\draw[white,fill=blue, ultra nearly transparent,rounded corners,draw=blue,line width=1.5pt] (-2,2.5) rectangle (6,5.5);

\draw[white,fill=red, ultra nearly transparent,rounded corners,draw=red,line width=1.5pt] (-2,2.5) rectangle (6,-0.5);
\draw[fill=yellow!20!white,draw=yellow,snake=coil,segment aspect=0] (-2,2.6) rectangle (6,2.4);

\node[white,circle,ball color=blue!60!green,draw=gray,line width=1.2pt,inner sep=1.5pt] (S11) at (0,4)  {$S$};
\node[white,circle,ball color=blue!60!green,draw=green!50!black,line width=1.2pt,inner sep=1.5pt] (S21) at (4,4)  {$S^*$};

\node[white,circle,ball color=blue!60!green,draw=gray,line width=1.2pt,inner sep=1.5pt] (S112) at (0,1)  {$S$};
\node[white,circle,ball color=blue!60!green,draw=green!50!black,line width=1.2pt,inner sep=1.5pt] (S212) at (4,1)  {$S^*$};
\node[white,rectangle,draw=white,fill=green!30!black,line width=1.2pt,inner sep=1.7pt] (X) at (2.5,3.3)  {$X$};
\node[white,rectangle,draw=white,fill=green!30!black,line width=1.2pt,inner sep=1.7pt] (Y) at (1.5,4.7)  {$Y$};

\node[white,rectangle,draw=white,fill=green!30!black,line width=1.2pt,inner sep=1.7pt] (X2) at (2.5,1.7)  {$X$};
\node[white,rectangle,draw=white,fill=green!30!black,line width=1.2pt,inner sep=1.7pt] (Y2) at (1.5,0.3)  {$Y$};

\draw[<->,draw=green!50!black,line width=2.5pt] (S11) .. controls (0.5,3.3) and (1.5,3.4) .. (X);
\draw[->,draw=green!50!black,line width=2.5pt] (X) .. controls (3,3.3) and (3.5,3.4) .. (S21);
\draw[<->,draw=gray,line width=2.5pt] (S21) .. controls (3.5,4.7) and (2.5,4.6) .. (Y);
\draw[->,draw=gray,line width=2.5pt] (Y) .. controls (1,4.7) and (0.5,4.6) .. (S11);

\draw[<->,draw=green!50!black,line width=2.5pt] (S112) .. controls (0.5,1.6) and (1.5,1.7) .. (X2);
\draw[->,draw=green!50!black,line width=2.5pt] (X2) .. controls (3,1.6) and (3.5,1.7) .. (S212);
\draw[<->,draw=gray,line width=2.5pt] (S212) .. controls (3.5,0.4) and (2.5,0.3) .. (Y2);
\draw[->,draw=gray,line width=2.5pt] (Y2) .. controls (1,0.4) and (0.5,0.3) .. (S112);

\node[diamond,draw=green!50!black,line width=1.2pt,fill=yellow!70!green,scale=0.6,inner sep=1.2pt] (E2) at (1,1.6)  {$E$};
\node[diamond,draw=gray,line width=1.2pt,fill=yellow!70!green,scale=0.6,inner sep=0.9pt] (F2) at (3,0.4)  {$F$};

\node[diamond,draw=green!50!black,line width=1.2pt,fill=yellow!70!green,scale=0.6,inner sep=1.2pt] (E) at (1,3.4)  {$E$};
\node[diamond,draw=gray,line width=1.2pt,fill=yellow!70!green,scale=0.6,inner sep=0.9pt] (F) at (3,4.6)  {$F$};

\draw[->,orange,line width=2pt] (S11.south west) -- node[black,draw=orange,line width=0.5pt,fill=white,inner sep=1pt,anchor=south] {$k_{19}$} (S112.north west);
\draw[->,orange,line width=2pt] (S112.north east) -- node[black,draw=orange,line width=0.5pt,fill=white,inner sep=1pt,anchor=north] {$k_{15}$} (S11.south east);
\draw[->,orange,line width=2pt] (E.south west) -- node[black,draw=orange,line width=0.5pt,fill=white,inner sep=1pt,anchor=south] {$k_{17}$} (E2.north west);
\draw[->,orange,line width=2pt] (E2.east) -- node[black,draw=orange,line width=0.5pt,fill=white,inner sep=1pt,anchor=north] {$k_{13}$} (E.east);
\draw[->,orange,line width=2pt] (X.south west) -- node[black,draw=orange,line width=0.5pt,fill=white,inner sep=1pt,anchor=south] {$k_{18}$} (X2.north west);
\draw[->,orange,line width=2pt] (X2.north east) -- node[black,draw=orange,line width=0.5pt,fill=white,inner sep=1pt,anchor=north] {$k_{14}$} (X.south east);
\draw[->,orange,line width=2pt] (S21.south west) -- node[black,draw=orange,line width=0.5pt,fill=white,inner sep=1pt,anchor=south] {$k_{20}$} (S212.north west);
\draw[->,orange,line width=2pt] (S212.north east) -- node[black,draw=orange,line width=0.5pt,fill=white,inner sep=1pt,anchor=north] {$k_{16}$} (S21.south east);
\end{tikzpicture}
\end{center}
\caption{Shuttling rates for the one-site phosphorylation cycle}\label{Fig:2}
\end{figure}

Specifically, we consider the polynomial  function $f_{\kappa}\colon\R^{12}\rightarrow \R^{12}$ given  by the right-hand side of the four conservation equations \eqref{totamounts} and  the equations in \eqref{odes} for all $\dot{x}_i$ except for  $\dot{x}_1,\dot{x}_2,\dot{x}_5$ and $\dot{x}_{11}$. The latter equations are  redundant and can be obtained from the conserved equations in \eqref{conseq}. The 12 components of the function $f_{\kappa}=(f_{\kappa,1},\ldots,f_{\kappa,12})$ are
{\small \begin{align*}
f_{\kappa,1} &= x_1+x_2+x_7+x_{11}, \\
 f_{\kappa,2} &=  x_2+x_3+x_4+x_6+x_8+x_9+x_{10}+x_{12}, \\ 
 f_{\kappa,3} &= x_5+x_6, \\
 f_{\kappa,4} &= x_{11}+x_{12}, \\
f_{\kappa,5} &= k_{2} x_{2} - k_{15} x_{3} - k_{1} x_{1} x_{3} + k_{6} x_{6} + k_{19} x_{9}, \\
 f_{\kappa,6} &=  k_{3} x_{2} - k_{16} x_{4} - k_{4} x_{4} x_{5} + k_{5} x_{6} + k_{20} x_{10},\\ 
f_{\kappa,7} &=  k_{4} x_{4} x_{5} - k_{5} x_{6} - k_{6} x_{6}, \\ 
f_{\kappa,8} &= k_{13} x_{1} - k_{17} x_{7} + k_{8} x_{8} + k_{9} x_{8} - k_{7} x_{7} x_{9},\\ 
 f_{\kappa,9} &=   k_{14} x_{2} - k_{8} x_{8} - k_{9} x_{8} - k_{18} x_{8} + k_{7} x_{7} x_{9}, \\ 
 f_{\kappa,10} &= k_{15} x_{3} + k_{8} x_{8} - k_{19} x_{9} - k_{7} x_{7} x_{9} +    k_{12} x_{12},\\ 
f_{\kappa,11} &=   k_{16} x_{4} + k_{9} x_{8} - k_{20} x_{10} - k_{10} x_{10} x_{11} +   k_{11} x_{12}, \\
f_{\kappa,12} &=  k_{10} x_{10} x_{11} - k_{11} x_{12} - k_{12} x_{12}.
\end{align*}}
If this function is injective over the real positive numbers $\R^n_+$, then multiple positive steady states with the same total amounts cannot occur.
As described in Section~\ref{heuristic}, we use the Jacobian injectivity criterion to investigate conditions on the rate constants for which the function is injective. Since $f_{\kappa}$ is quadratic, the criterion applies.
The determinant of the Jacobian of $f_{\kappa}$
can be computed using any software that enables algebraic (symbolic) computations, like Mathematica or Maple. We compute the determinant and extract the coefficients. These coefficients are polynomials in the rate constants and most of them contain only positive  summands. Therefore, we search for the coefficients that have negative summands. After appropriate factorization and simplification, we conclude that the coefficients are all positive if and only if the following expressions are positive:
{\small \begin{align*}
C_1 =& k_{9} k_{14}+k_{9} k_{17}+k_{3} (k_{18}-k_{17}) =k_{9} k_{14}+(k_{9}-k_3) k_{17}+k_{3}k_{18}, \\
C_2 =& k_{3} k_{12}(k_{15}-k_{16})(k_{18}-k_{17})+k_{3} k_{15} k_{16}(k_{18}-k_{17})  +k_{12} k_{15} k_{16} k_{18} + k_{9} k_{14} k_{15} k_{16} \\ & +k_{12} k_{14} k_{15} k_{16} +k_{12} k_{14} k_{16} k_{17} +k_{9} k_{15} k_{16} k_{17}+k_{12} k_{16} k_{17} k_{18}, \\
C_3 = & k_{3} k_{12} k_{15}( k_{18} -k_{17}) + k_{6} k_{9} k_{14} k_{15}+k_{6} k_{12} k_{14} k_{15}+k_{6} k_{12} k_{14} k_{17}+k_{6} k_{9} k_{15} k_{17}\\ & +k_{6} k_{12} k_{15} k_{18}+k_{6} k_{12} k_{17} k_{18}, \\
C_4=& k_{3} k_{15} (k_{18}-k_{17})k_{20}+k_{6} k_{9} k_{14} k_{15}+k_{6} k_{9} k_{15} k_{17}+k_{6} k_{9} k_{14} k_{20}+k_{6} k_{14} k_{15} k_{20} \\ &+ k_{9} k_{14} k_{15} k_{20}+k_{6} k_{9} k_{17} k_{20}+k_{6} k_{14} k_{17} k_{20}+k_{9} k_{15} k_{17} k_{20}+k_{6} k_{15} k_{18} k_{20}+k_{6} k_{17} k_{18} k_{20}, \\
C_5 =& k_{3} k_{13}+k_{9} (k_{14}-k_{13})+k_{3} k_{18} = (k_{3}-k_9) k_{13}+k_{9}k_{14}+k_{3} k_{18}, \\
C_6 =& k_{6} k_{9} (k_{14}-k_{13}) k_{19}+ k_{6} k_{12} k_{13} k_{14}+k_{6} k_{12} k_{13} k_{18}+k_{3} k_{12} k_{13} k_{19}+k_{6} k_{12} k_{14} k_{19} \\ & +k_{3} k_{12} k_{18} k_{19}+k_{6} k_{12} k_{18} k_{19}, \\
C_7 =& k_{9} (k_{14} -k_{13}) k_{16} k_{19}+k_{3} k_{12} k_{13} k_{16}+k_{12} k_{13} k_{14} k_{16}+k_{3} k_{12} k_{16} k_{18}+k_{12} k_{13} k_{16} k_{18}\\ & +k_{3} k_{12} k_{13} k_{19}+k_{3} k_{13} k_{16} k_{19}+k_{12} k_{14} k_{16} k_{19}+k_{3} k_{12} k_{18} k_{19}+k_{3} k_{16} k_{18} k_{19}+k_{12} k_{16} k_{18} k_{19}, \\
C_8 =& k_{6} k_{9}(k_{14}-k_{13})(k_{19}-k_{20}) +k_{9} (k_{14}-k_{13}) k_{19} k_{20} +k_{6} k_{13} k_{14} k_{20}+k_{6} k_{13} k_{18} k_{20} \\ & +k_{3} k_{13} k_{19} k_{20}+k_{6} k_{14} k_{19} k_{20}+k_{3} k_{18} k_{19} k_{20}+k_{6} k_{18} k_{19} k_{20}.
\end{align*}}
Observe that these expressions {\it only} involve  the 8 shuttling rates and $k_3,k_6,k_9,k_{12}$. Instances for which the coefficients $C_1,\dots,C_8$ are negative exist. If 
\begin{equation}\label{ineq}
k_{20}\leq k_{19},\quad  k_{18} \geq k_{17},\quad k_{16} \leq k_{15}, \quad k_{14}\geq k_{13},
\end{equation}
then $C_i>0$ for all $i$ and hence multistationarity cannot occur for any choice of total amounts (these inequalities correspond to Eqns.~(6) in the main text). However, there is no guarantee that when these conditions fail, the system admits multiple steady states for some total amounts. Figure~\ref{Fig:1} above is reproduced as Figure~\ref{Fig:2}  with  the shuttling rate constants  indicated.

\bigskip
We assume now that the dissociation constants are the same in the two compartments (the nucleus and the cytoplasm),    that is, we assume that $k_3=k_9$ and $k_6=k_{12}$.  In this case $C_i>0$  for all $i\neq 2,8$ and all shuttling rate constants. Therefore, two conditions suffice to guarantee monostationarity, namely:
\begin{align*}
\widetilde{C}_2= & k_{9} k_{12}(k_{15}-k_{16})(k_{18}-k_{17}) +
k_{9} k_{14} k_{15} k_{16}+k_{12} k_{14} k_{15} k_{16}\\ & +k_{12} k_{14} k_{16} k_{17} +k_{9} k_{15} k_{16} k_{18}+k_{12} k_{15} k_{16} k_{18}+k_{12} k_{16} k_{17} k_{18}>0, \\
\widetilde{C}_8 = & k_{9} k_{12} (k_{14}-k_{13})(k_{19}-k_{20})+k_{12} k_{13} k_{14} k_{20}+k_{12} k_{13} k_{18} k_{20}\\ & +k_{9} k_{14} k_{19} k_{20}+k_{12} k_{14} k_{19} k_{20}+k_{9} k_{18} k_{19} k_{20}+k_{12} k_{18} k_{19} k_{20}>0.
\end{align*}
The first corresponds to $C_2$ and the second to $C_8$.
By inspection of these two expressions, we conclude that multistationarity cannot occur  in any of the following cases:
\begin{align*}
\rm{(i)} & \quad k_{20}\leq k_{19}, \quad  k_{18} \geq k_{17},\quad k_{16} \leq k_{15}, \quad k_{14}\geq k_{13},\\
\rm{(ii)} & \quad k_{20}\geq k_{19},\quad  k_{18} \geq k_{17},\quad k_{16} \leq k_{15}, \quad k_{14}\leq k_{13}, \\
\rm{(iii)} & \quad k_{20}\leq k_{19},\quad  k_{18} \leq k_{17},\quad k_{16} \geq k_{15}, \quad k_{14}\geq k_{13}, \\
\rm{(iv)} & \quad k_{20}\geq k_{19},\quad   k_{18} \leq k_{17},\quad k_{16} \geq k_{15}, \quad k_{14}\leq k_{13}. 
\end{align*}
Note that these {\it only} involve the rate constants for the shuttling reactions. If the dissociation rate constants are not exactly the same in the cytoplasm and in the nucleus, but very similar, then the conditions above are still sufficient.

We see that the rate constants go in pairs: the shuttling rate constants of $S$ relate to those of $S^*$, and  the shuttling rate constants of $E$ to those of $X$. In particular, the following conditions are necessary for multistationarity:
\begin{enumerate}[(1)]
\item If $X$ shuttles into the nucleus slower than $E$ then $S$ shuttles into the cytoplasm slower than $S^*$ and vice versa.
\item If $X$ shuttles into the cytoplasm slower than $E$ then $S$ shuttles into the nucleus slower than $S^*$ and vice versa.
\end{enumerate}
Sets of rate constants for which $I_1<0$ can for instance  be obtained by letting the product $k_9k_{16}$ be large and the remaining products be small such that $k_{14} k_{15} + k_{12} k_{17} - k_{12} k_{18} + k_{15} k_{18}<0$ is satisfied.

\begin{table}[!h]
\small 
\centering
\begin{tabular}{|c|c|c|}
\hline
\rowcolor{lightgray}
$\#$ & No multistationarity &  Multistationarity   \\ \hline
1 & All & None \\ \hline
 \multirow{3}{*}{2} &  $\{S,S^*\}$ $\{E,Y\}$ $\{F,X\}$ $\{S^*,E\}$& $\{E,F\}$ $\{X,Y\}$ $\{S^*,X\}$ $\{S,Y\}$  \\ 
&   $\{S,F\}$ $\{S,X\}$ $\{S^*,Y\}$  $\{E,X\}$ & \\ 
&   $\{F,Y\}$ $\{S,E\}$ $\{S^*,F\}$ &
  \\ \hline
\multirow{3}{*}{3} &   $\{X,E,F\}$ $\{Y,E,F\}$ $\{X,Y,E\}$& $\{S,E,X\}$ $\{S^*,F,Y\}$  $\{S,E,Y\}$  $\{S^*,E,X\}$  \\ 
&  $\{X,Y,F\}$ $\{S,F,X\}$ $\{S^*,E,Y\}$ &  $\{S^*,F,X\}$ $\{S,E,S^*\}$ $\{S^*,F,S\}$  $\{S,S^*,X\}$ \\ &   $\{S,E,F\}$ $\{S^*,F,E\}$ & $\{S,X,Y\}$ $\{S^*,Y,X\}$ $\{S,F,Y\}$ $\{S,S^*,Y\}$     \\
\hline
\multirow{3}{*}{4}  & $\{Y,X,E,F \}$   & 
$\{S,S^*,X,F \}$ $\{S,S^*,Y,E \}$ $\{S,E,X,Y\}$ $\{S^*,F,X,Y \}$ \\ & &  $\{S,F,X,Y\}$ $\{S^*,E,X,Y \}$ $\{S,S^*,X,Y \}$ $\{S,S^*,E,F \}$  \\ & & 
$\{S,E,F,X\}$ $\{S^*,E,F,Y \}$ $\{S,E,F,Y\}$ \\ & &  $\{S^*,E,F,X \}$ $\{S,S^*,X,E \}$ $\{S,S^*,Y,F \}$ 
 \\  \hline 
5, 6 & None & All\\ \hline
\end{tabular}
\caption{One-site phosphorylation system. For all possible sets of shuttling species it is indicated if the system has the capacity for multiple steady states or not. }\label{Tab:1}
\end{table}

\medskip
\noindent
\textbf{Sets of shuttling species and multistationarity. }\label{creation}
We have shown that if the species $E,X,S,S^*$ shuttle between compartments, multistationarity is created. 
We next investigate what the sets of shuttling species that provide multistationarity are. 
The results are summarized in Table~\ref{Tab:1}. We use a systematic way  to classify each motif: First, we check if the system fulfills the Jacobian injectivity criterion for all rate constants. If the coefficients of the polynomial in $x$ given by the determinant of the Jacobian (as above) are all positive, then the system cannot exhibit multistationarity for any set of total amounts (see also \cite{Feliu-inj}).   If the criterion fails then we use the CRNT toolbox.

We have obtained that if only one species shuttles then multistationarity cannot occur. That is, at least two species, e.g. $\{S^*,X\}$ or $\{S,Y\}$, are required to create multistationarity in the one-site phosphorylation cycle for certain  total amounts and rate constants. The addition of shuttling species maintains multistationarity. 

\medskip
\noindent
 \textbf{Effects of varying the shuttling rates. }\label{bifur}
We analyze the steady-state response of $S^*$ in the nucleus as the shuttling rate constants and total amounts change in the system. 

The following table summarizes the type of saddle-node bifurcation curves obtained as shuttling rate constants of molecular species are varied.

\medskip
\begin{tabular}{|c|l|}
\hline
Rate constant & Rate-response curve \\ \hline 
$k_{13},k_{14},k_{19},k_{20}$ & For large rate constant, only a low stable steady state is obtained \\
$k_{15},k_{17},k_{18}$ &  For a small  rate constant, only a high stable steady state is obtained \\
$k_{16}$ & Similar to the previous case, but the high branch decreases (Fig.~\ref{f:bif1cyclek16}).   \\
\hline
\end{tabular}

\begin{figure}[!t]
\centering
\includegraphics[scale=0.5]{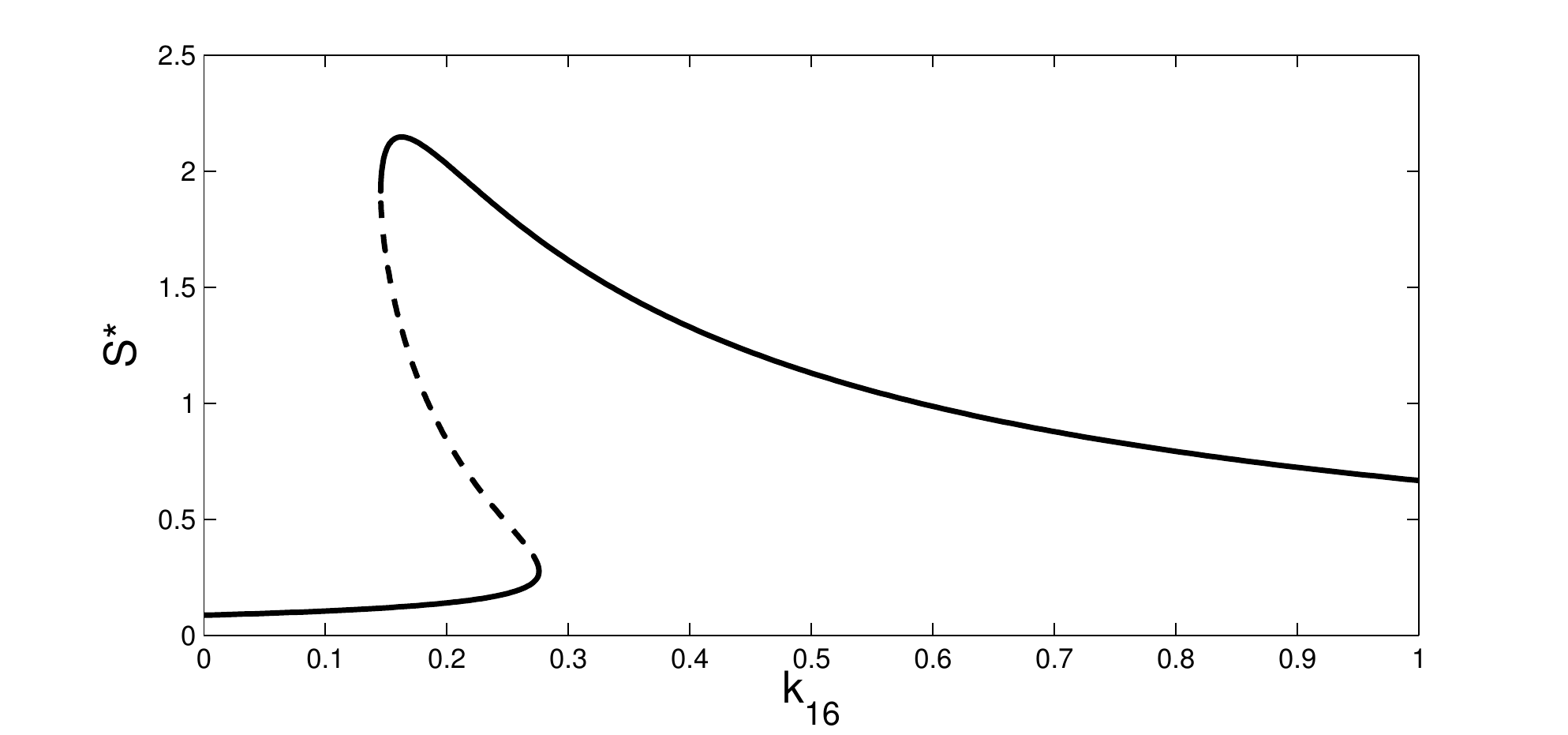}
\caption{The response $S^*$ plotted against the rate constant $k_{16}$, at baseline total amounts.} 
\label{f:bif1cyclek16} 
\end{figure}

\bigskip
By varying a total amount and shuttling rate constant simultaneously, the system may undergo irreversible switches. This occurs with respect to shuttling rate constants $k_{14},k_{15},k_{17},$ and $k_{20}.$
Specifically, for shuttling parameters, $k_{15}$ and $k_{17}$, the irreversible switch is obtained by either increasing $F_{tot}$ or decreasing $E_{tot}, S_{tot}$ or $F_{tot}^c$. As the value of the shuttling rate constant increases, the response curve switches from a low to a high steady state, favoring accumulation in the nucleus. Conversely, increasing the value of shuttling parameters $k_{14}$ and $k_{20}$ induces an irreversible switch from the high to low steady-state by either decreasing $F_{tot}$ or increasing $E_{tot}, S_{tot}$. As highlighted in the main text (see Figure 3), the $k_{20}$ bifurcation is irreversible at baseline parameter values.

\subsection{Shuttling in a two-site phosphorylation cycle}\label{twosite}
In eukaryotes, most protein phosphorylation events take place in more than one site. It is well known that multisite phosphorylation can cause multistationarity by itself \cite{Markevich-mapk,Thomson:2009fq}. However, multistationarity does not occur for all choices of rate constants.

We next investigate the effect of adding species compartmentalization in a two-site (sequential) phosphorylation system.
We first determine rate constants for which the two-site system cannot exhibit multistationarity. 
Then, we add species shuttling and  determine shuttling rate constants that induce multistationarity. 

\medskip
\noindent
\textbf{Conditions for monostationarity in a two-site phosphorylation cycle. }\label{preclusion2}
We consider a  two-site phosphorylation cycle in which modifications take place sequentially. The reactions describing the system are:

\centerline{
\xymatrix{
S_{0} + E \ar@<0.5ex>[r]^(0.6){k_1} & X_1  \ar@<0.5ex>[l]^(0.4){k_2} \ar[r]^(0.4){k_3} & S_{1} + E & 
S_{1} + E \ar@<0.5ex>[r]^(0.6){k_4} & X_2  \ar@<0.5ex>[l]^(0.4){k_5} \ar[r]^(0.4){k_6} & S_{2} + E \\
S_{1} + F \ar@<0.5ex>[r]^(0.6){k_7} & Y_1  \ar@<0.5ex>[l]^(0.4){k_8} \ar[r]^(0.4){k_9} & S_{0} + F &
S_{2} + F \ar@<0.5ex>[r]^(0.6){k_{10}} & Y_2  \ar@<0.5ex>[l]^(0.4){k_{11}} \ar[r]^(0.4){k_{12}} & S_{1} + F
}}

The set of species is ordered such that
$$(x_1,\dots,x_{9})=(E, X_1,S_0,S_1,F,Y_1,S_2,X_2,Y_2).$$
Assuming mass-action kinetics, then the differential equations describing the dynamics of the species concentrations  are:

\noindent
 \begin{align*}
\dot{x_1} &= k_{2} x_{2} + k_{3} x_{2} - k_{1} x_{1} x_{3} - k_{4} x_{1} x_{4} + k_{5} x_{8} +  k_{6} x_{8}, \\ 
 \dot{x_2} &= -k_{2} x_{2} - k_{3} x_{2} + k_{1} x_{1} x_{3}, \\
\dot{x_3} &= k_{2} x_{2} - k_{1} x_{1} x_{3} + k_{9} x_{6},  \\
\dot{x_4} &= k_{3} x_{2} - k_{4} x_{1} x_{4} - k_{7} x_{4} x_{5} + k_{8} x_{6} + k_{5} x_{8} +   k_{12} x_{9}, \\ 
\dot{x_5} &= -k_{7} x_{4} x_{5} + k_{8} x_{6} + k_{9} x_{6} -  k_{10} x_{5} x_{7} + k_{11} x_{9} + k_{12} x_{9}, \\
\dot{x_6} &= k_{7} x_{4} x_{5} - k_{8} x_{6} - k_{9} x_{6}, \\
\dot{x_7} &=  -k_{10} x_{5} x_{7} + k_{6} x_{8} +  k_{11} x_{9}, \\ 
 \dot{x_8} &= k_{4} x_{1} x_{4} - k_{5} x_{8} - k_{6} x_{8}, \\
\dot{x_9} &= k_{10} x_{5} x_{7} - k_{11} x_{9} - k_{12} x_{9}.
\end{align*}
This system has the following conserved amounts:
$$E_{tot} = x_1+x_2+x_8,\quad F_{tot}= x_5+x_6+x_9,\quad S_{tot} = x_2+x_3+x_4+x_6+x_7+x_8+x_9. $$
The steady-state equations are given by $\dot{x}_i=0$. Because of the constraints given by the conservation laws, the equations $\dot{x}_1=0$, $\dot{x}_2=0$ and $\dot{x}_5=0$ are redundant and can be removed. 

We proceed as above to determine rate constants for which the system cannot have multiple steady states. That is, we apply the Jacobian injectivity criterion. We consider  the function $f_{\kappa}\colon\R^{9}\rightarrow \R^{9}$ given  by the three equations coming from the conservation laws and the 6 remaining steady-state equations:
\begin{align*}
f_{\kappa,1}(x) & =x_1+x_2+x_8, \\
f_{\kappa,2}(x) & = x_2+x_3+x_4+x_6+x_7+x_8+x_9, \\
f_{\kappa,3}(x) & =x_5+x_6+x_{9}, \\
f_{\kappa,4}(x) & = k_{2} x_{2} - k_{1} x_{1} x_{3} + k_{9} x_{6},  \\
f_{\kappa,5}(x) & =k_{3} x_{2} - k_{4} x_{1} x_{4} - k_{7} x_{4} x_{5} + k_{8} x_{6} + k_{5} x_{8} +   k_{12} x_{9}, \\ 
f_{\kappa,6}(x) & =  k_{7} x_{4} x_{5} - k_{8} x_{6} - k_{9} x_{6}, \\
f_{\kappa,7}(x) & =   -k_{10} x_{5} x_{7} + k_{6} x_{8} +  k_{11} x_{9}, \\
f_{\kappa,8}(x) & = k_{4} x_{1} x_{4} - k_{5} x_{8} - k_{6} x_{8}, \\
f_{\kappa,9}(x) & = k_{10} x_{5} x_{7} - k_{11} x_{9} - k_{12} x_{9}.
\end{align*}
If this function is injective over the positive real numbers then multiple positive steady states cannot occur with the same total amounts in the conservation laws.
Next, we compute the determinant of the Jacobian of $f_{\kappa}$. As a polynomial in $x$, the only coefficients (which depend on the rate constants $k_i$) of the determinant that are not sums of positive terms are:
\begin{align*}
C_1 &= -k_{2} k_{4} k_{6} k_{7} k_{9} k_{10} - k_{3} k_{4} k_{6} k_{7} k_{9} k_{10} - 
 k_{1} k_{4} k_{6} k_{7} k_{9} k_{11} - k_{1} k_{4} k_{6} k_{7} k_{9} k_{12} \\ 
& +  k_{1} k_{3} k_{5} k_{7} k_{10} k_{12} + k_{1} k_{3} k_{6} k_{7} k_{10} k_{12} +  k_{1} k_{3} k_{4} k_{8} k_{10} k_{12} +  k_{1} k_{3} k_{4} k_{9} k_{10} k_{12} \\
 C_2 &=  -k_{1} k_{4} k_{7} k_{10} ( k_{6} k_{9}- k_{3} k_{12}). 
  \end{align*}
  If a choice of rate constants fulfills  $C_1,C_2>0$, then the system cannot have multiple positive steady states for any total amounts.
The coefficient $C_1$ can be rewritten as $$C_1 =   -k_{6} k_{9}k_{4} k_{7} (k_{2}  k_{10} + k_{3}   k_{10} + k_{1}   k_{11} + k_{1} k_{12})  + k_{3}k_{12}k_{1}k_{10}  (   k_{5} k_{7}  +   k_{6} k_{7}  +  k_{4} k_{8}   +   k_{4} k_{9} ).$$ 
 For $C_2>0$   we require 
 $$ k_{3}/k_{6} > k_{9}/k_{12}.  $$
  If this inequality is fulfilled then $C_1>0$ if also
$$  k_{4} k_{7} (k_{2}  k_{10} + k_{3}   k_{10} +
 k_{1}   k_{11} + k_{1} k_{12})   < k_{1}k_{10}  (   k_{5} k_{7}  +   k_{6} k_{7}  + 
 k_{4} k_{8}   +   k_{4} k_{9} ).$$ 
This inequality can be rewritten as
$$  \frac{k_{2}+k_{3}}{k_1} +
\frac{ k_{11} +  k_{12}}{k_{10}}   <       \frac{k_{5}+k_{6}}{k_{4}}+  \frac{k_{8}+k_{9}}{k_7}.  $$ 
Let
$$\alpha_1 = \frac{k_{5}+k_{6}}{k_{4}}- \frac{k_{2}+k_{3}}{k_1} , \qquad 
\alpha_2 =\frac{k_{8}+k_{9}}{k_7} -\frac{ k_{11} +  k_{12}}{k_{10}} , \qquad
\alpha_3 = \frac{k_{3}}{k_{6}} - \frac{k_{9}}{k_{12} }.
$$
If 
\begin{equation}\label{preclusion}
 \alpha_1,\alpha_2,\alpha_3>0,
\end{equation}
then the two-site phosphorylation system cannot have multiple positive steady states no matter the values of the total amounts. That is, a sufficient condition for the preclusion of multistationarity is obtained.

Observe that $\alpha_1$ and $\alpha_2$ imply an inequality between the Michaelis-Menten constants of the kinase and the phosphatase in each phosphorylation site. Namely, the Michaelis-Menten constant of $E$ for the second site is larger than that for the first  phosphorylation site, and the Michaelis-Menten constant of $F$ for the first site is larger than that for the second site.

Negation of this condition is a priori not  sufficient to guarantee multistationarity. However, if   $\alpha_3<0$, then there exist total amounts for which the motif exhibits multistationarity \cite{feliu:interface:2011}.

\medskip
\noindent
\textbf{Mass-action system of ordinary differential equations for the two-site shuttling. }\label{reactions2}
Consider now two copies of a two-site phosphorylation cycle as above and let $S_0,S_1,S_2,X_1,X_2,E$ shuttle between the nucleus and the cytoplasm. The reactions describing the system are:

\begin{itemize}
\item Reactions in the nucleus:

\centerline{
\xymatrix{
S_{0} + E \ar@<0.5ex>[r]^(0.6){k_1} & X_1  \ar@<0.5ex>[l]^(0.4){k_2} \ar[r]^(0.4){k_3} & S_{1} + E & 
S_{1} + E \ar@<0.5ex>[r]^(0.6){k_4} & X_2  \ar@<0.5ex>[l]^(0.4){k_5} \ar[r]^(0.4){k_6} & S_{2} + E \\
S_{1} + F \ar@<0.5ex>[r]^(0.6){k_7} & Y_1  \ar@<0.5ex>[l]^(0.4){k_8} \ar[r]^(0.4){k_9} & S_{0} + F &
S_{2} + F \ar@<0.5ex>[r]^(0.6){k_{10}} & Y_2  \ar@<0.5ex>[l]^(0.4){k_{11}} \ar[r]^(0.4){k_{12}} & S_{1} + F
}}

\item Reactions in the cytoplasm:

\centerline{
\xymatrix{
S_{0}^c + E^c \ar@<0.5ex>[r]^(0.6){k_{13}} & X_1^c  \ar@<0.5ex>[l]^(0.4){k_{14}} \ar[r]^(0.4){k_{15}} & S_{1}^c + E^c & 
S_{1}^c + E^c \ar@<0.5ex>[r]^(0.6){k_{16}} & X_2^c  \ar@<0.5ex>[l]^(0.4){k_{17}} \ar[r]^(0.4){k_{18}} & S_{2}^c + E^c \\
S_{1}^c + F^c \ar@<0.5ex>[r]^(0.6){k_{19}} & Y_1^c  \ar@<0.5ex>[l]^(0.4){k_{20}} \ar[r]^(0.4){k_{21}} & S_{0}^c + F^c &
S_{2}^c + F^c \ar@<0.5ex>[r]^(0.6){k_{22}} & Y_2^c  \ar@<0.5ex>[l]^(0.4){k_{23}} \ar[r]^(0.4){k_{24}} & S_{1}^c + F^c
}}

\item Shuttling reactions:

 \centerline{\xymatrix{
E  \ar@<0.3ex>[r]^(.5){k_{25}}  & E^c  \ar@<0.3ex>[l]^(.5){k_{31}}  & 
 X_1  \ar@<0.3ex>[r]^(.5){k_{26}}  & X_1^c  \ar@<0.3ex>[l]^(.5){k_{32}} &
  S_0  \ar@<0.3ex>[r]^(.5){k_{27}}  & S_0^c  \ar@<0.3ex>[l]^(.5){k_{33}}  \\
  S_1  \ar@<0.3ex>[r]^(.5){k_{28}}  & S_1^{c}  \ar@<0.3ex>[l]^(.5){k_{34}}   &
    S_2  \ar@<0.3ex>[r]^(.5){k_{29}}  & S_2^c  \ar@<0.3ex>[l]^(.5){k_{35}}   &
 X_2  \ar@<0.3ex>[r]^(.5){k_{30}}  & X_2^c  \ar@<0.3ex>[l]^(.5){k_{36}}   
}}
\end{itemize}

The species are ordered as:
$$(x_1,\dots,x_{9})=(E, X_1,S_0,S_1,F,Y_1,S_2,X_2,Y_2)$$
$$(x_{10},\dots,x_{18})=(E^c, X_1^c,S_0^c,S_1^{c},F^c,Y_1^c,S_2^c,X_2^c,Y_2^c).$$
Assuming mass-action kinetics, the system of differential equations describing the dynamics of the concentrations of the species is:

\noindent
 \begin{align*}
\dot{x}_1 &= k_{2} x_{2} + k_{3} x_{2} - k_{1} x_{1} x_{3} - k_{4} x_{1} x_{4} + k_{5} x_{8} +  k_{6} x_{8}-k_{25}x_1+k_{31}x_{10}, \\ 
 \dot{x}_2 &= -k_{2} x_{2} - k_{3} x_{2} + k_{1} x_{1} x_{3}-k_{26}x_2+k_{32}x_{11}, \\
\dot{x}_3 &= k_{2} x_{2} - k_{1} x_{1} x_{3} + k_{9} x_{6}-k_{27}x_3+k_{33}x_{12},  \\
\dot{x}_4 &= k_{3} x_{2} - k_{4} x_{1} x_{4} - k_{7} x_{4} x_{5} + k_{8} x_{6} + k_{5} x_{8} +   k_{12} x_{9}-k_{28}x_4+k_{34}x_{13}, \\ 
\dot{x}_5 &= -k_{7} x_{4} x_{5} + k_{8} x_{6} + k_{9} x_{6} -  k_{10} x_{5} x_{7} + k_{11} x_{9} + k_{12} x_{9}, \\
\dot{x}_6 &= k_{7} x_{4} x_{5} - k_{8} x_{6} - k_{9} x_{6}, \\
\dot{x}_7 &=  -k_{10} x_{5} x_{7} + k_{6} x_{8} +  k_{11} x_{9}-k_{29}x_7+k_{35}x_{16}, \\ 
 \dot{x}_8 &= k_{4} x_{1} x_{4} - k_{5} x_{8} - k_{6} x_{8}-k_{30}x_8+k_{36}x_{17}, \\
\dot{x}_9 &= k_{10} x_{5} x_{7} - k_{11} x_{9} - k_{12} x_{9}, \\
\dot{x}_{10} &= k_{14} x_{11} + k_{15} x_{11} - k_{13} x_{10} x_{12} - k_{16} x_{10} x_{13} + k_{17} x_{17} +  k_{18} x_{17}+k_{25}x_1-k_{31}x_{10}, \\ 
 \dot{x}_{11} &= -k_{14} x_{11} - k_{15} x_{11} + k_{13} x_{10} x_{12}+k_{26}x_2-k_{32}x_{11}, \\
\dot{x}_{12} &= k_{14} x_{11} - k_{13} x_{10} x_{12} + k_{21} x_{15}+k_{27}x_3-k_{33}x_{12},  \\
\dot{x}_{13} &= k_{15} x_{11} - k_{16} x_{10} x_{13} - k_{19} x_{13} x_{14} + k_{20} x_{15} + k_{17} x_{17} +   k_{24} x_{18}+k_{28}x_4-k_{34}x_{13}, \\ 
\dot{x}_{14} &= -k_{19} x_{13} x_{14} + k_{20} x_{15} + k_{21} x_{15} -  k_{22} x_{14} x_{16} + k_{23} x_{18} + k_{24} x_{18}, \\
\dot{x}_{15} &= k_{19} x_{13} x_{14} - k_{20} x_{15} - k_{21} x_{15}, \\
\dot{x}_{16} &=  -k_{22} x_{14} x_{16} + k_{18} x_{17} +  k_{23} x_{18}+k_{29}x_7-k_{35}x_{16}, \\ 
 \dot{x}_{17} &= k_{16} x_{10} x_{13} - k_{17} x_{17} - k_{18} x_{17}+k_{30}x_8-k_{36}x_{17}, \\
\dot{x}_{18} &= k_{22} x_{14} x_{16} - k_{23} x_{18} - k_{24} x_{18}.
\end{align*}
This system has the following conserved amounts:
\begin{align*}
E_{tot} &=   x_1+x_2+x_8+x_{10}+x_{11}+x_{17}, \\
  F_{tot} &= x_5+x_6+x_9, \\
  F_{tot}^c &= x_{14}+x_{15}+x_{18}, \\ 
S_{tot} &= x_2+x_3+x_4+x_6+x_7+x_8+x_9+x_{11}+x_{12}+x_{13}+x_{15}+x_{16}+x_{17}+x_{18}.
 \end{align*}

\medskip
\noindent
\textbf{Creation of multistationarity in a two-site phosphorylation cycle. }\label{multi}
We use the CRNT toolbox to obtain initial rate constants and total amounts for which the system admits multiple positive steady states. The output rate constants do not fulfill  $\alpha_1,\alpha_2,\alpha_3>0$ in each compartment independently. Hence, it is not possible to decide whether multistationarity arises due to shuttling or due to phosphorylation of two different sites. 

Next we investigate the effect of changing the  rate constants with respect to the existence of multistationarity for the same total amounts. We proceed by manually modifying the rates while keeping multistationarity and such that the sufficient conditions for the preclusion of multistationarity in a two-site phosphorylation cycle are satisfied in each compartment. 

We end up with the following rate constants:
\begin{itemize}
\item Reaction rates in the nucleus:
\begin{align*}
k_1 &=100 & k_2&= 2 & k_3&= 10 & k_4&=80 & k_5&= 6 & k_6&= 6  \\ 
k_7&= 350 & k_8&= 3 & k_9&= 10 & k_{10}&= 650 & k_{11}&= 8 & k_{12}&= 8. 
\end{align*}
\item Reaction rates in the cytoplasm:
\begin{align*}
k_{13} &=300  & k_{14}&=1 & k_{15}&= 10 & k_{16}&= 50 & k_{17}&= 1 & k_{18}&= 1 \\ 
k_{19}&= 350 & k_{20}&=30 & k_{21}&= 190 & k_{22}&= 150 & k_{23}&= 2 & k_{24}&= 20.
\end{align*}
\item Shuttling rates:
\begin{align*}
k_{25}&=10 & k_{26} &=  30 & k_{27} &=  70 & k_{28} &=  30 & k_{29} &=  1 & k_{30} &=  10 \\ k_{31} &=  450 & k_{32} &=  20 & k_{33} &=  20 & k_{34} &=  25 & k_{35} &=  10 & k_{36} &=  100.
\end{align*}
\end{itemize}
This choice of rate constants fulfills that $\alpha_1,\alpha_2,\alpha_3>0$ in the nucleus and in the cytoplasm (where in the later, indices of the rate constants in $\alpha_*$ are shifted by $12$). Therefore, with these rate constants, the two-site phosphorylation cycles  in the nucleus and in the cytoplasm cannot have multiple positive steady states independently of each other.

The system with the shuttling reactions, however, does have the capacity for multistationarity. Specifically, 
if the total amounts are set to:
$$E_{tot} = 50,\qquad S_{tot} = 100 \qquad F_{tot} = 15\qquad 
F_{tot}^c=21,$$ 
then the system has three positive steady states: two of them are stable and one is unstable. 
The positive steady states are the following:
{\small \begin{align*}
SS_1 =& (1.89, 9.18, 0.62, 1.88, 0.01, 0.7, 25.11, 23.21, 14.28,  0.07, 13.38,  6.8, 0.04, 18.49, 1.15, 0.01, 2.28, 1.36) \\ 
  SS_2 =& (6.45, 5.93, 0.1, 0.61, 0.01, 0.17, 35.4, 25.64, 14.82,  0.13,  9.33, 2.8, 0.03, 18.32, 0.8, 0.02, 2.52,  1.89) \\
  SS_3 =& (0.37, 17.59, 5.96, 2.3, 0.17, 10.66, 0.6, 5.65, 4.16, 0.04, 25.8, 24.89, 0.06, 19.22, 1.72, 0.00044, 0.56, 0.06).
  \end{align*}}
Here $SS_1$ is the unstable steady state.

\begin{figure}[!t]
\centering
\includegraphics[scale=0.4]{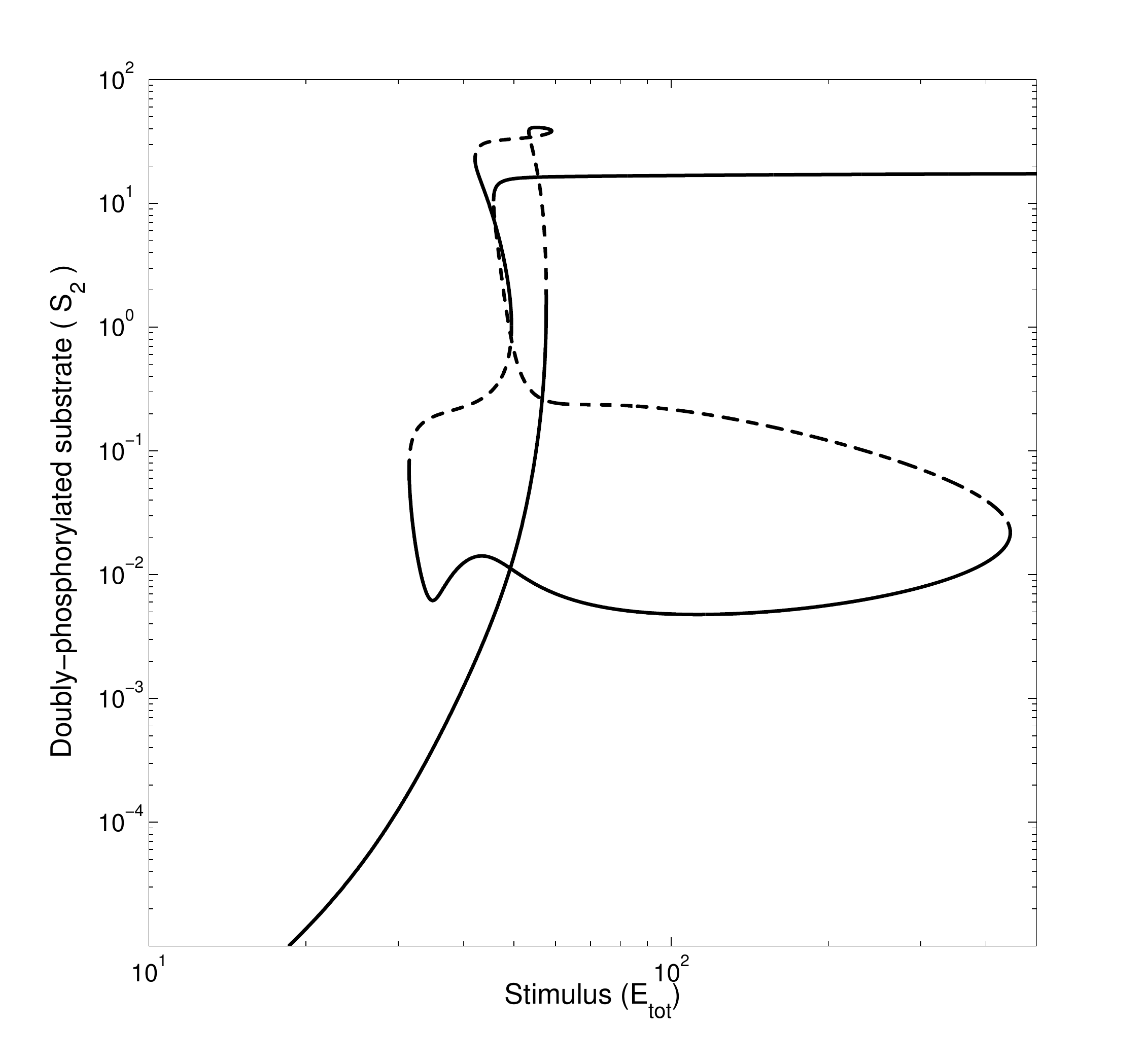}
\caption{The steady state response curve of $S_2$ as stimulus ($E_{tot}$) varies. All other parameters are at baseline values. Stable state, solid line; unstable state, dashed line.} 
\label{f:loglog7statesEtotvsS2} 
\end{figure}

\medskip
\noindent
\textbf{Rate constants to obtain seven steady states (for Figs. 4(B-D) in the main text). }\label{7ss}
Additionally, we have observed that, with specific choices of rate constants and total amounts, up to 7 steady states can be created in this system. For instance, consider the rate constants (used in the main text to create Figures 4(B-D)):
\begin{align*}
(k_1,\dots,k_{12}) &=  (101, 2, 11, 79, 6, 6, 568, 3, 12, 1502, 8, 8), \\
 (k_{13},\dots,k_{24}) &= (210, 3, 34, 49, 1, 1, 344, 26, 187, 149, 1, 1),\\
(k_{25},\dots,k_{36}) &=  (10, 34, 7.5, 312.5, 0.1, 0.1, 44, 23, 2, 250, 0.1, 0.1),
\end{align*}
and the total amounts $(E_{tot},S_{tot},F_{tot},F_{tot}^c)=(57, 111, 15, 21)$.
This system has 7 steady states, 4 of which are stable:
{\small \begin{align*}
SS_1 &= (18.27, 2.41, 0.04, 0.03, 5.05, 6.28, 0.01,4.12, 3.67, 5.2, 1.45, 0.005,0.21, 0.01,0.002, 46.2, 25.61, 20.99) \\
  SS_2 &= (12.66, 1.2, 0.03, 0.17, 0.47, 3.11, 0.26, 14.67, 11.42, 3.4,0.73, 0.004, 0.3, 0.01, 0.004, 33.77, 24.34, 20.99) \\
  SS_3 &= (13.01, 0.05, 0.0009, 0.23, 0.01, 0.09, 16.34, 19.9, 14.9, 2.96, 0.06, 0.003, 0.29, 0.02, 0.01, 17.14, 21.06, 20.1) \\
  SS_4 &= (13.66, 2.04, 0.004, 0.23, 0.004, 0.039, 35.84, 20.54, 14.96,  2.63, 3.95, 0.3, 0.22, 2.48, 0.84, 0.1, 14.18, 17.76) \\
  SS_5 &= (14.56, 3.53, 0.006, 0.22, 0.004, 0.03, 40.64, 20.63, 14.96, 2.49, 6.85, 0.56, 0.14, 6.54, 1.45, 0.03, 8.95, 13.01) \\
  SS_6 &= (4, 15.03, 0.4,   0.49, 0.27, 5.07, 0.38, 12.88, 9.65, 0.12, 23.78, 35.81, 0.2, 14.89, 4.88, 0.001, 1.19,1.23) \\ 
  SS_7 &= (5.94, 11.14, 0.15, 0.5, 0.02,   0.43, 6.89, 19.51, 14.55, 0.12,  18.91, 31.22, 0.17, 14.86, 4.07, 0.002, 1.38, 2.07)
  \end{align*}}
 The stable steady states are $SS_1,SS_3,SS_5,SS_6$.
 
\bigskip 
As $E_{tot}$ varies, the number of states changes as shown on a log-log scale in Figure~\ref{f:loglog7statesEtotvsS2}, corresponding to the bifurcation diagram (Figure 4B in the main text, semi-log scale).

\section{Software tools}
Calculations to assess multistationarity were made using Mathematica (Wolfram Research, Inc., Mathematica, Version 7.0, Champaign, IL., 2010) and CRNT toolbox \cite{CRNT-toolbox}. Bifurcation diagrams were computed using Oscill8 \cite{oscill8} and visualized with MATLAB (The MathWorks Inc., Natick, MA, R2011b). Figures in the main text were created using Adobe Illustrator (Versions CS3 and CS4. San Jose, California: Adobe Systems, Inc., 2011).

\end{document}